\documentclass[preprintnumbers,prd,twocolumn,showpacs,floatfix,preprintnumbers,superscriptaddress,nofootinbib]{revtex4-2}

\usepackage{graphicx}
\usepackage{epsfig}
\usepackage{bm}
\usepackage{amssymb}
\usepackage{float}
\usepackage{amsmath}
\usepackage{subfigure}
\usepackage{dcolumn}
\usepackage[colorlinks]{hyperref}
\usepackage{cleveref}
\usepackage[usenames,dvipsnames]{color}
\hypersetup{
     breaklinks=true,
    pdfstartview={FitH},    
    colorlinks=true,       
    linkcolor=blue,          
    citecolor=red,        
    filecolor=magenta,      
    urlcolor=magenta,           
    anchorcolor=green,      
    linktocpage=true
}
\usepackage{orcidlink}

\newcommand{\Mpl}{M_{\textrm{Pl}}}

\newcommand{\nn}{\nonumber}

\def\doi{http://doi.org}

\allowdisplaybreaks

\begin{document}

\title{Gaussian process reconstruction of scalar field dynamics from recent cosmological data}

\author{Afaq Maqsood\orcidlink{0009-0000-3084-9169}}
\email{afaq.res@gmail.com}
\affiliation{Department of Physics, Jamia Millia Islamia, New Delhi, 110025, India}
\author{Sonej Alam\orcidlink{0009-0008-8322-2923}}
\email{sonejalam36@gmail.com}
\author{Md. Wali Hossain\orcidlink{0000-0001-6969-8716}}
\email{ mhossain@jmi.ac.in}
\affiliation{Department of Physics, Jamia Millia Islamia, New Delhi, 110025, India}


\begin{abstract}
We reconstruct the late-time expansion history and dark energy dynamics using available cosmological data. We consider dark energy as an effective minimally coupled canonical scalar field without assuming a specific form for its potential. For reconstruction, we use Gaussian Process (GP) regression with a joint dataset consisting of 32 cosmic chronometer measurements (CC32), DESI DR2 baryon acoustic oscillation data, three Type Ia supernova compilations (Pantheon+, Union3, and DES Y5), and compressed CMB distance priors. From the reconstructed Hubble parameter and its derivatives, we obtain the scalar field kinetic and potential energy densities, the equation of state $w(z)$, and the dimensionless slope parameter $|\lambda(z)|$ of the scalar field potential. The reconstructed potential is nearly constant at late times, with $\tilde{V}(0) \simeq 0.68$--$0.70\rho_{c,0}$, close to the dark-energy density in flat $\Lambda$CDM. The kinetic term remains small over $0 \lesssim z \lesssim 2.5$, while $w(z)$ remains close to $-1$ within the uncertainties. Using only CC32 and DESI DR2, we find a mild ($1\sigma$) hint of a phantom-divide crossing near $z\sim0.5$. However, this feature depends on the supernova compilation and cannot be regarded as a firm conclusion with current data. The slope parameter $|\lambda(z)|$ shows mild evolution, with present-day central values around $0.8$--$1.0$, but large uncertainties due to its dependence on higher-order derivatives of the expansion history. The curvature parameter $\Gamma(z)$ requires even higher-order derivatives and is not constrained by current data; therefore, we do not report its reconstruction. Allowing small spatial curvature, $\Omega_k=0$--$0.02$, changes the results only slightly and remains within the existing uncertainty bands.
\end{abstract}

\maketitle

\flushbottom

\section{Introduction}
\label{sec:intro}

The discovery that the Universe is undergoing accelerated expansion \cite{SupernovaSearchTeam:1998fmf,SupernovaCosmologyProject:1998vns} is one of the central milestones of modern cosmology. Observations from type Ia supernovae first provided compelling evidence for this phenomenon, and the recent availability of high-precision supernova datasets further strengthens this foundation. In particular, modern compilations such as Pantheon+ \cite{Scolnic:2017caz,Brout:2022vxf}, Union~3.1 \cite{Rubin:2023jdq}, and DES~Y5 \cite{DES:2024jxu}, consisting of large samples of luminosity-distance measurements extending to redshifts $z \sim 2$, provide some of the most comprehensive mappings of the late time expansion history. This supernova evidence has since been complemented by independent cosmological probes, including cosmic microwave background (CMB) radiation \cite{Planck:2013pxb,Planck:2018vyg}, baryon acoustic oscillations (BAO) \cite{DESI:2024kob,DESI:2025zgx,10.1093/mnras/,eBOSS:2018cab,2017MNRAS.470.2617A,2017A&A...608A.130D,DESI:2024mwx,DESI:2024uvr,DESI:2024lzq,DESI:2024aqx}, and cosmic chronometers (CC) \cite{Zhang_2014, Stern:2010cv,Moresco:2012jh,Moresco_2016,10.1093/mnras/stx301,10.1093/mnrasl/slv037,Moresco:2020fbm}, which provide information about the Hubble parameter across a broad redshift range. Most recently, BAO measurements from the Dark Energy Spectroscopic Instrument (DESI), including its first data release (DR1) and second data release (DR2), have significantly enhanced constraints at intermediate redshifts. Taken together, these datasets constitute a precise mapping of the late time expansion history to date and form the empirical backbone of modern cosmology.

Within the standard cosmological model, acceleration is attributed to a cosmological constant, leading to the flat $\Lambda$CDM framework. This simple model has been remarkably successful in explaining a wide range of observations. Nevertheless, it faces several important challenges. Among these are the well-known fine-tuning \cite{Martin:2012bt} and coincidence problems \cite{Zlatev:1998tr,Steinhardt:1999nw} associated with the cosmological constant. Observationally, tensions have emerged in recent years, most notably the discrepancy between local measurements and early-universe measurements of the Hubble constant \cite{Planck:2013pxb,Planck:2018vyg,Riess:2021jrx,Krolewski:2025deb,Li:2025mmo,Kamionkowski:2022pkx}, and possible inconsistencies in the growth of cosmic structures. These difficulties motivate the exploration of alternative explanations for cosmic acceleration, including dynamical dark energy models \cite{Copeland:2006wr,Sahni:1999gb,Hossain:2025grx,Colgain:2024mtg,Capozziello:2025lor,Shlivko:2026jxa,Li:2026hwq,You:2025uon,Ormondroyd:2025iaf}.

A prominent class of such alternatives is scalar field dark energy \cite{Ratra:1987rm,Wetterich:1987fm,Hossain:2025grx,Sohail:2024oki,Ramadan:2024kmn,Yao:2025wlx,Gialamas:2025pwv}, where the late time acceleration \cite{Copeland:2006wr} is driven by the interplay between the potential and kinetic energy of a scalar field \cite{Nair:2013sna}. The slope and curvature of the scalar potential plays a central role in determining the dynamics of the field and, consequently, the background expansion history of the Universe  \cite{Hossain:2023lxs,Hossain:2025grx}. Traditionally, analyses of scalar field models adopt specific parameterizations\cite{Chevallier:2000qy,Linder:2002et,Efstathiou:1999tm,Jassal:2005qc,Akthar:2024tua,Alam:2025epg,Gokcen:2026pkq} for the equation of state or for the potential itself. While useful, such parametric approaches can lead to biased conclusions if the assumed forms are not sufficiently general. This has motivated the development of reconstruction techniques, which aim to recover cosmological functions directly from data in a model-independent manner.  

Among non-parametric approaches, Gaussian Process (GP)  regression\cite{Rasmussen:2005gp,Holsclaw:2010sk,Holsclaw:2010nb,Holsclaw:2011wi} has emerged as a powerful tool. GP is a  Bayesian nonparametric statistical framework that treats functions as random variables characterized by a mean and a covariance kernel. Given observational data, GPs provide smooth reconstructions of cosmological quantities \cite{Busti:2014aoa,Busti:2014dua,Seikel:2012uu,Seikel:2012cs,Seikel:2013fda,MAQSOOD2026140456,Shafieloo:2012ht,Mukherjee:2022yyq,Mukherjee:2020vkx,Ruchika:2025mkx,Velazquez:2024aya,Mukherjee:2024ryz,Favale:2023lnp,Jiang:2025ilh,Favale:2025mgk,Jiang:2024xnu,Johnson:2025blf,Ormondroyd:2025exu,Sangwan:2017kxi,Niu:2023hak,You:2025uon,GuptaChoudhury:2026gsl,Maqsood:2026krg} and their derivatives without requiring a prescribed functional form. This flexibility makes the method particularly well suited for studying the dark energy equation of state $w(z)$ \cite{Holsclaw:2010sk,Holsclaw:2010nb}, scalar field potentials, and other derived quantities. Previous works have successfully employed GP techniques to reconstruct $H(z)$ \cite{MAQSOOD2026140456,Mukherjee:2022yyq,Mukherjee:2020vkx,Ruchika:2025mkx,Velazquez:2024aya,Jiang:2025ilh,Jesus:2021bxq,Niu:2023hak}, the equation of state parameter $w(z)$, and scalar field dynamics from various datasets, demonstrating the robustness of this approach.  

In this work, we employ GP to reconstruct the Hubble parameter $H(z)$ and its derivatives using a joint dataset consisting of cosmic chronometers, DESI DR2 BAO measurements, Type Ia supernova compilations (Pantheon+, Union3, and DES Y5), together with CMB distance priors. From these reconstructions, we obtain model independent estimates of several key cosmological quantities, including the dark energy equation of state $w(z)$, the scalar field components, namely the potential energy $V(z)$\cite{Nair:2013sna,Jesus:2021bxq} and kinetic term $K(z)$\cite{Nair:2013sna}, as well as the dimensionless slope parameter $\lambda(z)$. The curvature parameter $\Gamma(z)$ involves higher-order derivatives of the reconstructed quantities, for which the resulting reconstruction was found to be less reliable and is therefore not presented in this work. By avoiding any assumed parameterization and relying solely on observational data\cite{Scolnic:2017caz,Brout:2022vxf,Rubin:2023jdq,DES:2024jxu,Planck:2013pxb,Planck:2018vyg,DESI:2024kob,DESI:2025zgx,10.1093/mnras/,eBOSS:2018cab,2017MNRAS.470.2617A,2017A&A...608A.130D,DESI:2024mwx,DESI:2024uvr,DESI:2024lzq,DESI:2024aqx,Zhang_2014,Stern:2010cv,Moresco:2012jh,Moresco_2016,10.1093/mnras/stx301,10.1093/mnrasl/slv037,Moresco:2020fbm}, this framework allows us to test whether the inferred evolution is compatible with $\Lambda$CDM or exhibits signatures of dynamical dark energy\cite{Copeland:2006wr,Sahni:1999gb,Hossain:2025grx,Colgain:2024mtg,Capozziello:2025lor,Shlivko:2026jxa,Li:2026hwq,You:2025uon,Ormondroyd:2025iaf}. We further assess the complementary impact of DESI DR2 measurements~\cite{DESI:2025zgx}, Type Ia supernova compilations\cite{Scolnic:2017caz,Brout:2022vxf,Rubin:2023jdq,DES:2024jxu}, and CMB distance priors\cite{Chen:2018dbv}, whose progressive inclusion systematically tightens constraints and improves the robustness of the reconstruction.

The paper is organized as follows. In the next section, we present the reconstruction methodology. In Sect.~\ref{sec:GP}, we describe how GP are used in this work. Sec.~\ref{sec:data} outlines the observational datasets employed in this work. In Sec.~\ref{sec:rec}, we present the reconstruction of the expansion history, including the reconstruction of the Hubble parameter in Sec.~\ref{sec:H}, followed by the reconstruction of the scalar-field potential in Sec.~\ref{sec:V} and kinetic term in Sec.~\ref{sec:K}. We then reconstruct the dimensionless slope parameter $\lambda(z)$ in Sec.~\ref{sec:lambda} and the equation of state $w(z)$ in Sec.~\ref{sec:w}. We summarize our findings and conclude in Sec.~\ref{sec:discussion}.

\section{Methodology}

Gaussian Processes provides a non-parametric framework for reconstructing a continuous function directly from observational data, along with its associated uncertainties \cite{Seikel:2012uu,Seikel:2012cs,Seikel:2013fda,Shafieloo:2012ht,Mukherjee:2022yyq,Mukherjee:2020vkx,Ruchika:2025mkx,Velazquez:2024aya,Mukherjee:2024ryz}. A detailed description of this method is presented in Sec.~\ref{sec:GP}. In this work, we employ GP to reconstruct the Hubble parameter $H(z)$ and its derivatives \cite{Seikel:2012cs,Seikel:2013fda,Shafieloo:2012ht,Mukherjee:2022yyq,MAQSOOD2026140456,Mukherjee:2020vkx,Ruchika:2025mkx,Velazquez:2024aya,Mukherjee:2024ryz,Favale:2023lnp,Favale:2025mgk,Li:2025ops,Koksbang:2026wvh,Jesus:2021bxq,Jesus:2022xwb,Liu:2023agr}. A key advantage of this approach is that derivatives of a Gaussian Process are themselves Gaussian Processes, enabling a statistically robust reconstruction of $H'(z)$ and higher-order derivatives without introducing numerical noise.

The expansion history $H(z)$ is reconstructed directly from observational data in a model-independent manner, without assuming a specific spatial geometry. The background spacetime is described by a Friedmann--Lema\^{i}tre--Robertson--Walker (FLRW) metric, while spatial curvature is introduced only at the level of derived quantities.

The dynamics of the Universe are governed by the Friedmann equations,
\begin{align}
3 \left(H^2+\frac{ k}{a^2}\right) \Mpl^2 &= \rho_{\rm total}, \\
\left(2\dot H +3 H^2+\frac{k}{a^2}\right) \Mpl^2 &= -p_{\rm total},
\end{align}
where {\it dot} represents derivative with respect to (w.r.t.) time $t$. As we focus on late time cosmology, we assume that the total energy content of the Universe consists of pressureless matter and a dark energy component, such that $\rho_{\rm total} = \rho_{\rm m} + \rho_{\rm de}$ and $p_{\rm total} = p_{\rm de}$, where the subscripts ${\rm m}$ and ${\rm de}$ denote matter and dark energy, respectively. 

Without assuming any specific parametrization for the expansion history or the dark energy equation of state, we perform a non-parametric reconstruction directly from observational data. Nevertheless, to interpret the reconstructed dark energy sector within a physically motivated framework, we consider an effective dynamical dark energy description in which the energy density and pressure are expressed in terms of kinetic and potential contributions. Such a decomposition is naturally motivated by scalar field models of dynamical dark energy \cite{Copeland:2006wr}. Accordingly, we write
\begin{equation}
\rho_{\rm de} = K + V, \qquad p_{\rm de} = K - V,
\end{equation}
where $K(z)$ and $V(z)$ denote the effective kinetic and potential energy densities of the dark energy component, respectively.
 Using the relation $d/dt = -H(z)(1+z)\,d/dz$, these quantities can be expressed directly in terms of the reconstructed expansion history. It is convenient to normalize all energy densities by the present critical density $\rho_{c0} = 3H_0^2\Mpl^2$, and define
\begin{equation}
\tilde{K}(z) = \frac{K(z)}{\rho_{c0}}, \qquad
\tilde{V}(z) = \frac{V(z)}{\rho_{c0}}.
\end{equation}
Throughout this paper, we normalize all the observables to critical density for reconstruction processes.  

Solving the Friedmann equations, we obtain
\begin{equation}
\label{eq:K}
\tilde{K}(z) = \frac{(1+z)}{3H_0^2} H H' - \frac{\Omega_m}{2}(1+z)^3 - \frac{\Omega_k}{3}(1+z)^2,
\end{equation}
\begin{equation}
\label{eq:V}
\tilde{V}(z) = \frac{H^2}{H_0^2} - \frac{(1+z)}{3H_0^2} H H' - \frac{\Omega_m}{2}(1+z)^3 - \frac{2\Omega_k}{3}(1+z)^2.
\end{equation}

These expressions show that the  kinetic and potential contributions are fully determined by the expansion history $H(z)$ and its first derivative. The equation of state parameter is then given by
\begin{equation}
w(z) = \frac{p_{\rm de}}{\rho_{de}} = \frac{\tilde{K}(z) - \tilde{V}(z)}{\tilde{K}(z) + \tilde{V}(z)}.
\label{eq:w}
\end{equation}
Thus, once $H(z)$ and its derivatives are reconstructed using GP, all relevant dark energy quantities follow directly in a model-independent manner. 

Within this framework, it is useful to define the dimensionless slope and curvature parameters of the potential, given by
\begin{equation}
\lambda = -\frac{V_{,\phi}}{V}, \qquad 
\Gamma= \frac{V_{,\phi \phi}\,V}{V_{,\phi}^2},
\label{eq:lam+gamma}
\end{equation}
where subscript $\phi$ represents derivative with respect to $\phi$. $\lambda$ and $\Gamma$ characterize the steepness and curvature of the effective potential and play a central role in determining the dynamical behavior of scalar field models\cite{Ratra:1987rm,Wetterich:1987fm,Hossain:2025grx,Sohail:2024oki,Ramadan:2024kmn,Yao:2025wlx,Gialamas:2025pwv,Hossain:2023lxs,Hossain:2025grx}.

Using the relation between derivatives with respect to $\phi$ and redshift, these expressions can be transformed into functions of observable quantities. In particular, we obtain
\begin{equation}
\lambda(z)
=
\frac{1}{V(z)}\,\frac{dV}{dz}\,H(z)(1+z)\,\frac{1}{\dot{\phi}(z)},
\label{eq:lambda}
\end{equation}

while the curvature parameter becomes
\begin{align}
%
\Gamma &= \frac{(1+z)H(z)}{\lambda(z)^2 \dot{\phi}^2 V(\phi)} \left[ \frac{d}{dz} \left( (1+z)H(z) \frac{dV}{dz} \right) - 3H(z) \frac{dV}{dz} \right] \nn \\ &+ \frac{(1+z)^2 H(z)^2}{\lambda(z)^2 \dot{\phi}^4 V(\phi)} \left( \frac{dV}{dz} \right)^2
\label{eq:Gamma}
\end{align}
In particular, $\lambda(z)$ quantifies the slope of the potential and governs the rolling behavior of the scalar field, while $\Gamma(z)$ encodes information about its curvature, characterizing the nature of the potential. The reconstruction of $\Gamma(z)$ is considerably more challenging, since its definition in Eq.~\eqref{eq:Gamma} involves higher-order derivatives of the reconstructed Hubble parameter and consequently leads to substantially larger uncertainties; we therefore do not reconstruct $\Gamma(z)$ in this work. In contrast, although $|\lambda(z)|$ can be reconstructed, its uncertainty is also strongly amplified by its dependence on higher-order derivatives. We therefore use $|\lambda(z)|$ and characterize its uncertainty through a Monte Carlo sampling of the joint GP posterior, allowing the non-Gaussian distribution of the reconstructed slope to be properly captured.

To account for the non-Gaussian uncertainty in the nonlinear quantity
$|\lambda(z)|$, we sample the joint Gaussian Process posterior of
$[H(z),H'(z),H''(z)]$ and evaluate $|\lambda(z)|$ for each realization. For a
set of $N$ Monte Carlo realizations $\{|\lambda_i(z)|\}_{i=1}^{N}$, the
central value is taken as the median,
\begin{equation}
|\lambda(z)|_{\mathrm{med}} =
\mathrm{median}\left(\{|\lambda_i(z)|\}_{i=1}^{N}\right),
\label{eq:lam_med1}
\end{equation}
while the lower and upper bounds are defined by the 16th and 84th percentiles
of the same sample,
\begin{equation}
\begin{aligned}
|\lambda(z)|_{\mathrm{low}} &= Q_{0.16}\left(\{|\lambda_i(z)|\}_{i=1}^{N}\right),\\
|\lambda(z)|_{\mathrm{high}} &= Q_{0.84}\left(\{|\lambda_i(z)|\}_{i=1}^{N}\right),
\end{aligned}
\label{eq:lam_med2}
\end{equation}
where $Q_{p}$ denotes the $p$-th quantile of the sample. Thus, the uncertainty
band is given by $[|\lambda|_{\mathrm{low}},|\lambda|_{\mathrm{high}}]$,
corresponding to the central $68\%$ credible interval of the Monte Carlo
posterior~\cite{Gelman2013,Barua:2025jid}.

\section{Gaussian Processes}
\label{sec:GP}
A Gaussian Process (GP) is a nonparametric framework used to reconstruct an unknown function directly from data without assuming a specific parametric form\cite{Rasmussen:2005gp,Holsclaw:2010sk,Holsclaw:2010nb,Holsclaw:2011wi}. In this approach, the function values are treated as jointly Gaussian-distributed random variables characterized by a mean function and a covariance kernel.

A GP is defined as
\begin{equation}
f(x) \sim GP(\mu(x), k(x, x')),
\end{equation}
where $\mu(x)$ is the mean function and $k(x,x')$ is the covariance kernel, which encodes the correlation between function values and determines the smoothness of the reconstruction.

Given observational data $y$ at points $X=\{x_i\}$, the GP predicts the function at new points through a Gaussian distribution with mean and covariance determined by the kernel and the data covariance. The kernel hyperparameters are optimized by maximizing the logarithmic marginal likelihood
\begin{equation}
\ln P(y) = -\frac12 (y-\mu)^{T} [K+C]^{-1} (y-\mu)
-\frac12 \ln |K+C|
-a,
\end{equation}
where $K$ is the covariance matrix, $a=\frac{n}{2} \ln (2\pi)$ and $C$ represents observational uncertainties. 

In this analysis, we employ a zero mean function, $\mu(z)=0$ \cite{Mukherjee:2022yyq,Mukherjee:2020vkx,Ruchika:2025mkx,Johnson:2025blf,MAQSOOD2026140456}, a commonly adopted choice in cosmological GP reconstructions that reduces prior bias and allows the expansion history to be predominantly determined by the data. We have also checked that using alternative simple mean functions, such as a constant mean, leads to statistically consistent reconstructions. This indicates that the inferred results are largely insensitive to the specific choice of mean function and are driven primarily by the observational data rather than by prior assumptions. A key property of GP is that derivatives of a GP are themselves GP's\cite{Rasmussen:2005gp,Holsclaw:2010sk}. This allows one to reconstruct not only the function but also its derivatives in a statistically consistent manner. Consequently, cosmological quantities depending on $f$, $f'$, and higher derivatives can be directly obtained from the GP without introducing numerical differentiation noise.

The choice of covariance kernel in GP encodes prior assumptions about the smoothness and differentiability of the reconstructed function. While the Squared Exponential kernel leads to infinitely differentiable functions and often results in over-smoothing, the Mat\'ern family provides a more flexible framework. The general Mat\'ern kernel is given by
\begin{equation}
k_{\nu}(d) = \sigma^2 \frac{2^{1-\nu}}{\Gamma(\nu)} \left( \frac{\sqrt{2\nu}d}{\ell} \right)^{\nu} K_{\nu}\left( \frac{\sqrt{2\nu}d}{\ell} \right)
\end{equation}
where $\nu$ controls the smoothness of the function. In this work, we adopt the Mat\'ern $7/2$ kernel, which ensures that the reconstructed function is sufficiently smooth and differentiable up to higher-order derivatives required for cosmological analysis, while still retaining enough flexibility to capture genuine features in the data. Lower-order kernels such as Mat\'ern $3/2$ and $5/2$ are inadequate due to insufficient differentiability, whereas higher-order choices approach the overly smooth behavior of the Squared Exponential kernel. Therefore, the Mat\'ern $7/2$ kernel provides an optimal balance between smoothness and flexibility for reconstructing $H(z)$ and its derivatives. 

It is worth noting that, based on the logarithm of marginal likelihood (LML) and leave-out-one cross validation (LOO-CV) method \cite{MAQSOOD2026140456}, no particular kernel among Matern family and Squared Exponential(RBF) is strongly preferred by the data, instead the primary distinction between kernels arises from their smoothness properties, governed by the parameter $\nu$. Consequently, the choice of kernel is guided by the level of differentiability required for reconstructing higher-order cosmological quantities, rather than by a significant statistical preference in LML  \cite{Johnson:2025blf,MAQSOOD2026140456}.

\section{Data}
\label{sec:data}
\subsection{Cosmic Chronometer $H(z)$ Data}
Cosmic chronometers provide a direct model-independent measurement of the Hubble parameter $H(z)$ from the differential age evolution of passively evolving galaxies. The expansion rate is given by
\begin{equation}
  H(z) = -\frac{1}{1+z}\frac{dz}{dt},
\end{equation}
where $dz/dt$ is estimated from galaxy populations at nearby redshifts. This method assumes a FLRW spacetime and passive stellar evolution calibrated via a specific initial mass function and star formation history, but does not require a specific cosmological model for the background expansion.

We use a compilation of 32 cosmic chronometer measurements (CC32), spanning the redshift range $0.07 < z < 1.965$ \cite{Zhang_2014, Stern:2010cv,Moresco:2012jh,Moresco_2016,10.1093/mnras/stx301,10.1093/mnrasl/slv037,Moresco:2020fbm}. 
The uncertainties are incorporated into the Gaussian Process reconstruction through the covariance matrix,
\begin{equation}
C_{ij} = C^{\mathrm{stat}}_{ij} + C^{\mathrm{sys}}_{ij},
\end{equation}
where the statistical component is diagonal, while the systematic contribution introduces correlations due to effects such as stellar population modeling and metallicity \cite{Moresco:2020fbm}. This leads to a non-diagonal covariance matrix, ensuring consistent propagation of uncertainties into the reconstructed $H(z)$ and its derivatives. 
\subsection{DESI DR2 BAO data}
We incorporate baryon acoustic oscillation (BAO) measurements from the second data release of the Dark Energy Spectroscopic Instrument (DESI DR2) \cite{DESI:2025zgx}, which provide precise constraints on the expansion history over intermediate and high redshifts. The DESI DR2 catalog reports measurements in terms of $D_H(z)/r_d$, $D_M(z)/r_d$, and $D_V(z)/r_d$, where $r_d$ denotes the sound horizon at the drag epoch.

For Gaussian Process reconstruction, it is advantageous to use observables directly related to the Hubble expansion rate. Among the reported BAO quantities, the radial distance scale
\begin{equation}
\frac{D_H(z)}{r_d} \equiv \frac{c}{H(z)\, r_d}
\end{equation}
provides a direct probe of $H(z)$ at discrete redshifts. The transverse comoving distance is given by
\begin{equation}
D_M(z)=c\int_0^z \frac{dz'}{H(z')},
\end{equation}
while the volume-averaged distance
\begin{equation}
D_V(z)=\left[z\,D_M^2(z)\,D_H(z)\right]^{1/3}
\end{equation}
represents a compressed combination of radial and transverse BAO information. To avoid introducing redundant information and possible covariance double counting, we do not include the $D_V(z)/r_d$ measurements in the present analysis.

We therefore use the DESI DR2 measurements of $D_H(z)/r_d$, together with the corresponding $D_M(z)/r_d$ constraints, obtained from multiple tracers including luminous red galaxies (LRG), emission line galaxies (ELG), quasars (QSO), and the Lyman-$\alpha$ forest, spanning effective redshifts $z_{\rm eff}=0.510,\,0.706,\,0.934,\,1.321,\,1.484,$ and $2.330$. To convert the radial BAO measurements into Hubble parameter values, we adopt the sound horizon scale $r_d = 147.05 \pm 0.30~{\rm Mpc}$ from the Planck 2018 analysis \cite{Planck:2018vyg}. We note that this choice introduces a mild dependence on early-Universe physics. However, recent analyses such as \cite{Zhang:2025bmk} adopt a nearly identical value of $r_d \simeq 147~{\rm Mpc}$. The Hubble parameter is then obtained as
\begin{equation}
H(z) = \frac{c}{(D_H/r_d)\, r_d}.
\end{equation}
The associated uncertainty in $H(z)$ is computed through standard error propagation, while correlations between BAO measurements are incorporated using the covariance matrix provided by the DESI collaboration. 

\subsection{Type Ia supernovae}
We use three Type~Ia supernova datasets: Pantheon+ 
\cite{Scolnic:2017caz,Brout:2022vxf}, Union3 \cite{Rubin:2023jdq}, and 
DES~Y5 \cite{DES:2024jxu}. The Pantheon+ sample consists of 1701 light curves corresponding to 1550 spectroscopically confirmed supernovae, spanning the redshift range $0.001 \lesssim z \lesssim 2.26$, with improved calibration and reduced systematic uncertainties compared to earlier compilations. The primary observable is the distance modulus,
\begin{equation}
\mu(z) = 5 \log_{10}\!\left(\frac{d_L(z)}{\mathrm{Mpc}}\right) + 25,
\end{equation}
From this, the dimensionless comoving distance can be constructed and used in the reconstruction of cosmological quantities.

In this work, we use the Pantheon+ compilation\cite{Scolnic:2017caz,Brout:2022vxf}, which consists of 1550 spectroscopically confirmed SNe Ia covering the redshift range $0.001 < z < 2.26$. This dataset provides high-quality measurements of the distance modulus along with a full covariance matrix that includes both statistical and systematic uncertainties. Compared to earlier compilations, Pantheon+ offers improved precision and better control over systematics, making it well suited for model-independent analyses.

To examine the stability of our reconstruction, we further include two independent Type Ia supernova compilations. The Union3 dataset~\cite{Rubin:2023jdq} consists of a carefully calibrated sample of SNe Ia with reliable distance measurements that have been extensively employed in cosmological analyses. We also make use of the DES Year 5 (DES Y5) supernova sample~\cite{DES:2024jxu}, which provides high-quality observations from the Dark Energy Survey, offering extended redshift coverage and improved photometric calibration.

Each supernova compilation is incorporated using its measured distance moduli together with the corresponding covariance matrix, accounting for both statistical and systematic uncertainties. As standard candles, Type Ia supernovae constrain the integrated expansion history through the luminosity distance rather than the Hubble parameter directly. Consequently, when combined with direct expansion probes such as cosmic chronometers and DESI BAO measurements, they substantially strengthen the reconstruction by yielding tighter constraints, especially over the low- and intermediate-redshift regime.

\subsection{CMB data}
We incorporate Cosmic Microwave Background (CMB) information through a compressed likelihood approach based on the shift parameters $R$ and $l_a$, derived from the Planck 2018 data \cite{Planck:2018vyg} and are obtained by Chen et al  \cite{Chen:2018dbv}. These quantities capture the dominant geometrical information from the CMB and provide an efficient summary of early Universe constraints.

In this analysis, the CMB data are not included as inputs to the Gaussian Process (GP) reconstruction; instead, they are used as a post-reconstruction consistency test of the inferred expansion history.

The CMB observables are defined as
\begin{equation}
R = \sqrt{\Omega_m} \frac{H_0 D_A(z_*)}{c}, \qquad 
l_a = \pi \frac{D_A(z_*)}{r_s(z_*)},
\end{equation}
where $D_A(z_*)$ is the angular diameter distance to the decoupling redshift $z_*$. The quantities $z_*$ and $r_s(z_*)$ are evaluated using standard fitting formulas.

For the CMB sector, we adopt fixed cosmological parameters consistent with Planck 2018, namely $H_0 = 67.2 \, \mathrm{km\,s^{-1}\,Mpc^{-1}}$, $\Omega_m = 0.315$, and $\Omega_b = 0.049$\cite{Planck:2018vyg}. These parameters are used only for the evaluation of CMB observables and do not affect the GP reconstruction.

The theoretical predictions are compared with the observed Planck values using the corresponding $2\times2$ covariance matrix,
$\chi^2_{\mathrm{CMB}} = \Delta F_{\mathrm{CMB}}^{\mathrm{T}} C_{\mathrm{CMB}}^{-1} \Delta F_{\mathrm{CMB}}$,
where $\Delta F_{\mathrm{CMB}} = (R - R_{\mathrm{obs}},\, l_a - l_a^{\mathrm{obs}})$. This approach allows us to test the consistency between the reconstructed late time expansion history and early-Universe constraints encoded in the CMB.
The values $\Omega_k = 0$, $0.01$, and $0.02$ are chosen to probe mild departures from spatial flatness while remaining close to the nearly flat geometry favored by current cosmological observations~\cite{Planck:2018vyg,DESI:2024kob,DESI:2025zgx}.
The reconstruction is performed using six combinations of observational datasets:
\begin{itemize}\setlength\itemsep{0em}

    \item[(i)] CC32
    
    \item[(ii)] CC32 + DESI DR2
    
    \item[(iii)] CC32 + DESI DR2 + Pantheon+
    
    \item[(iv)] CC32 + DESI DR2 + Pantheon+ + CMB
    
    \item[(v)] CC32 + DESI DR2 + Union3 + CMB
    
    \item[(vi)] CC32 + DESI DR2 + DES Y5 + CMB

\end{itemize}

These combinations are employed to assess the robustness of the reconstruction across different late time probes. Each dataset plays a complementary role in constraining the expansion history. Within the Gaussian Process framework, all datasets are incorporated consistently in terms of the Hubble parameter. While cosmic chronometers and BAO data directly constrain $H(z)$, supernova observations anchor the distance evolution, ensuring a smooth and self-consistent reconstruction of the expansion history and its derivatives \cite{Seikel:2013fda,Shafieloo:2012ht,Mukherjee:2022yyq,Mukherjee:2020vkx,Ruchika:2025mkx,Velazquez:2024aya,Mukherjee:2024ryz,Favale:2023lnp,Favale:2025mgk,Li:2025ops,Koksbang:2026wvh}. The CMB data are incorporated in the form of compressed shift parameters\cite{Chen:2018dbv} and primarily act as a high-redshift consistency check, without directly driving the reconstruction.

It is important to emphasize that the inclusion of the supernova data modifies the covariance structure of the joint Gaussian Process reconstruction, since the kernel hyperparameters are optimized simultaneously over all datasets. As nearly $95\%$ of the supernova measurements and most CC data lie at $z<1$, the reconstruction is tightly constrained at low redshift. Although DESI DR2 provides precise measurements over $0.5\lesssim z\lesssim1.4$, with only one point near $z\approx2.3$, the decreasing data density toward higher redshift broadens the GP covariance and increases the uncertainties in the reconstructed expansion history. These uncertainties are progressively amplified in quantities involving higher-order derivatives of $H(z)$, from $V(z)$, $K(z)$, and $w(z)$ to $|\lambda(z)|$, which depends on $H''(z)$ (equivalently $D'''(z)$), and ultimately to $\Gamma(z)$, which requires even higher-order derivatives.

\begin{figure*}[t]
    \centering

    \includegraphics[width=0.32\textwidth]{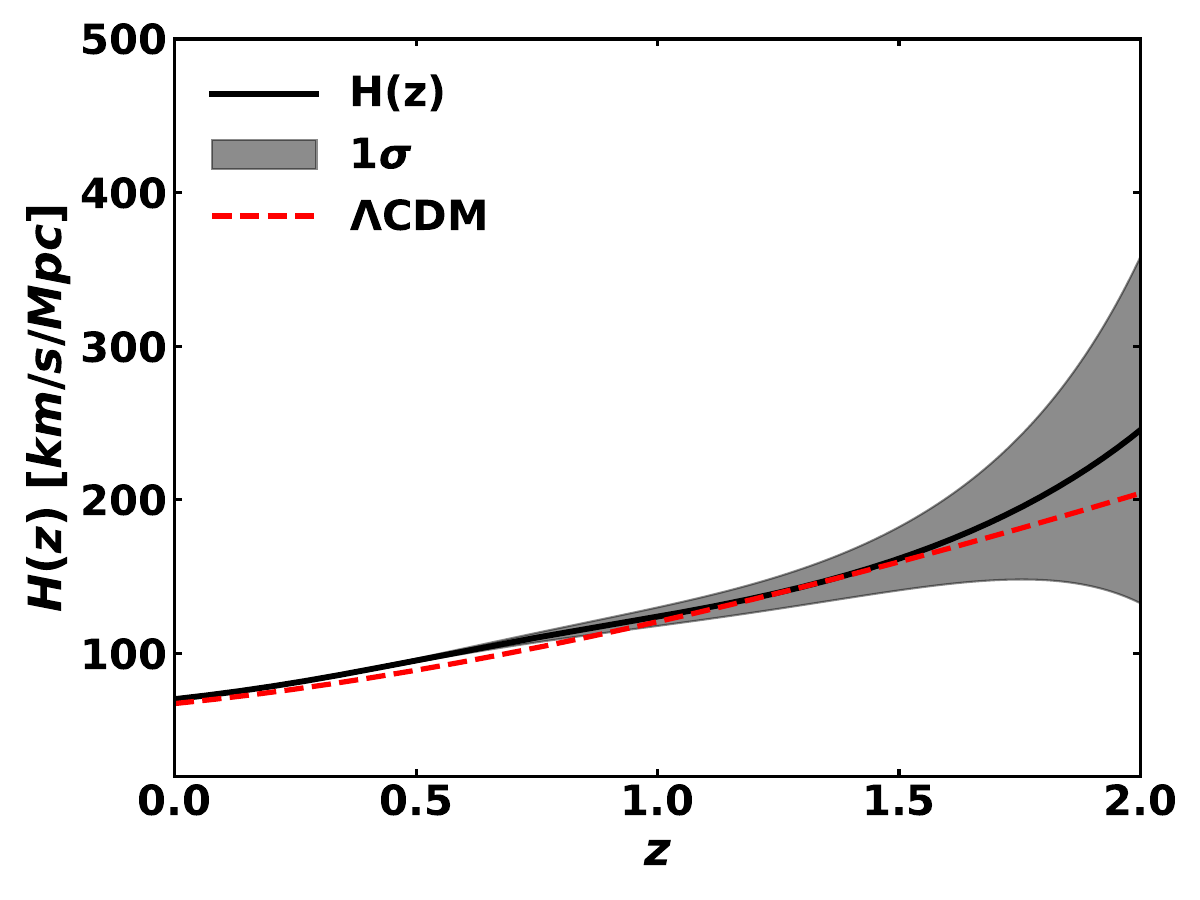}
    \hfill
    \includegraphics[width=0.32\textwidth]{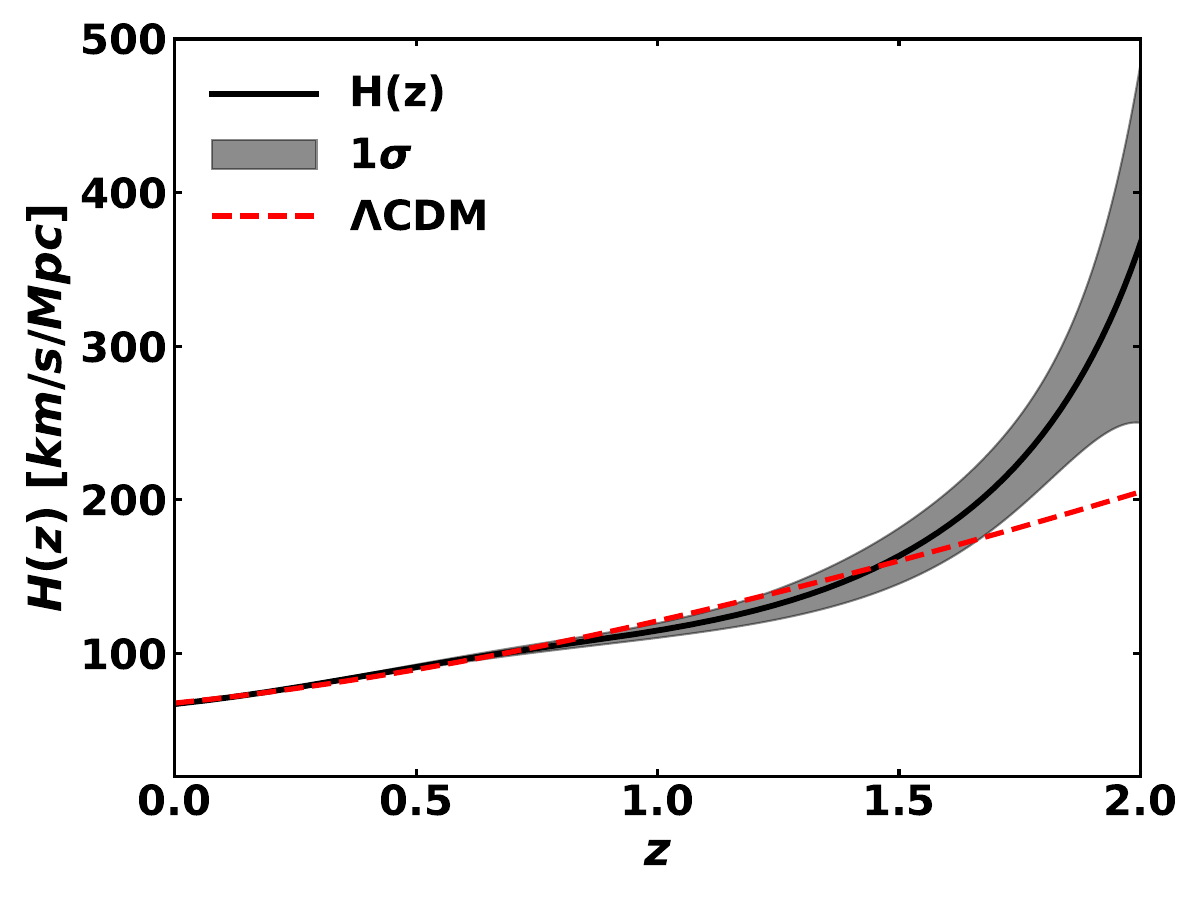}
    \hfill
    \includegraphics[width=0.32\textwidth]{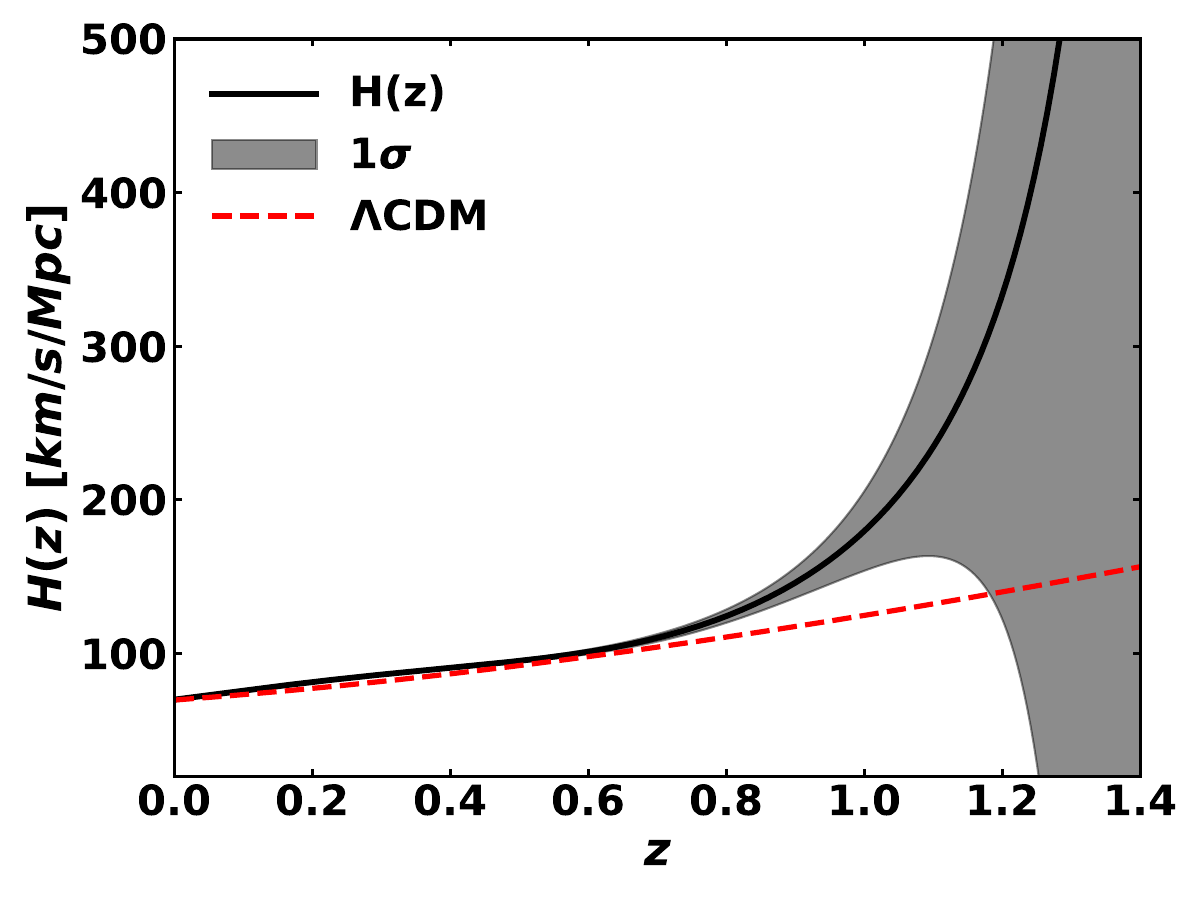}

    \caption{Gaussian Process reconstruction of the Hubble parameter $H(z)$ using different combinations of datasets. The left panel shows CC32 + DESI DR2 + Pantheon+ + CMB, the middle panel shows CC32 + DESI DR2 + Union3 + CMB, and the right panel shows CC32 + DESI DR2 + DES Y5 + CMB. The solid line represents the GP reconstruction, while the shaded regions indicate the $1\sigma$ confidence intervals.}
    \label{fig:H}
\end{figure*}
\section{Reconstruction of scalar field dynamics}
\label{sec:rec}
\subsection{Reconstruction of Hubble Parameter}
\label{sec:H}

In this section, we reconstruct the Hubble parameter $H(z)$ using the GP method. The reconstruction is based on a combination of complementary datasets. The Cosmic Chronometer (CC32) data provide direct, model-independent measurements of the expansion rate, forming the backbone of the reconstruction. The inclusion of DESI DR2 BAO measurements improves the precision by anchoring the reconstruction at intermediate redshifts with high accuracy. The addition of CMB information further constrains the global behavior of $H(z)$ by providing a high-redshift anchor and fixing the overall normalization. The inclusion of different late time datasets mainly affects the low-redshift region, where a noticeable tightening of the uncertainty bands is observed due to their strong statistical constraining power. Minor variations between the three cases reflect differences in the statistical weight and redshift coverage of the respective datasets, but do not alter the overall behavior of the expansion history.

The reconstructed $H(z)$ and its derivatives \cite{Seikel:2012cs,Seikel:2013fda,Shafieloo:2012ht,Mukherjee:2022yyq,MAQSOOD2026140456,Mukherjee:2020vkx,Ruchika:2025mkx,Velazquez:2024aya,Mukherjee:2024ryz,Favale:2023lnp,Favale:2025mgk,Li:2025ops,Koksbang:2026wvh,Jesus:2021bxq,Jesus:2022xwb,Liu:2023agr} serve as the fundamental inputs for all subsequent analyses presented in this work. These quantities are used consistently to derive further cosmological observables in a fully model-independent manner \cite{Seikel:2012cs,Seikel:2013fda,Shafieloo:2012ht,Mukherjee:2022yyq,Mukherjee:2020vkx,Ruchika:2025mkx,Velazquez:2024aya,Mukherjee:2024ryz,Favale:2023lnp}. It is important to note that the GP reconstruction of $H(z)$ itself does not assume any specific spatial geometry. The effect of spatial curvature enters only at the level of derived quantities through the corresponding cosmological equations (~\ref{eq:K},~\ref{eq:V},~\ref{eq:lambda},~\ref{eq:Gamma}), allowing us to explicitly assess its impact without affecting the data-driven reconstruction. 

\subsection{Reconstruction of the Scalar Field Potential}
\label{sec:V} 

\begin{figure*}[t]
    \centering

    \includegraphics[width=0.48\textwidth]{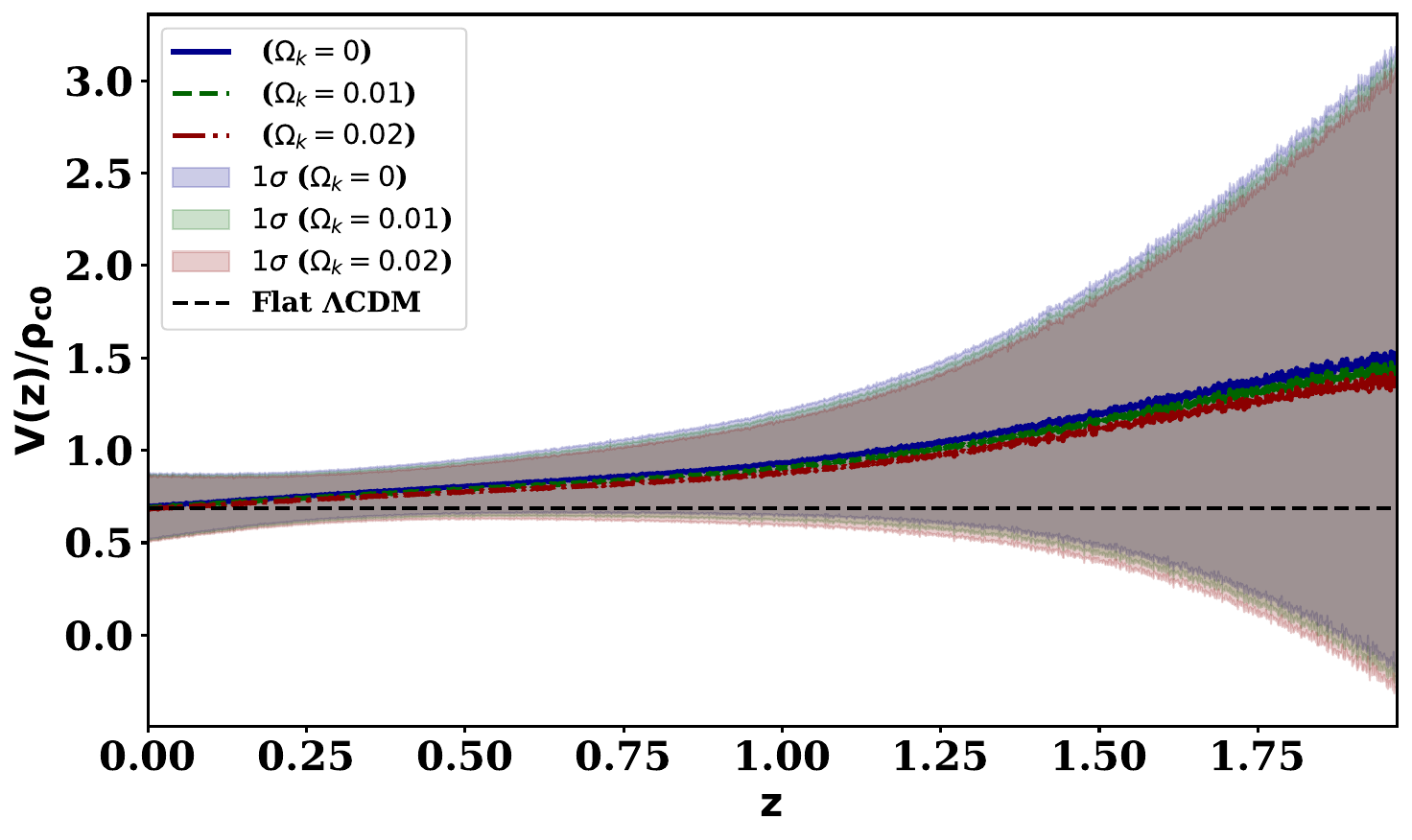}
    \includegraphics[width=0.48\textwidth]{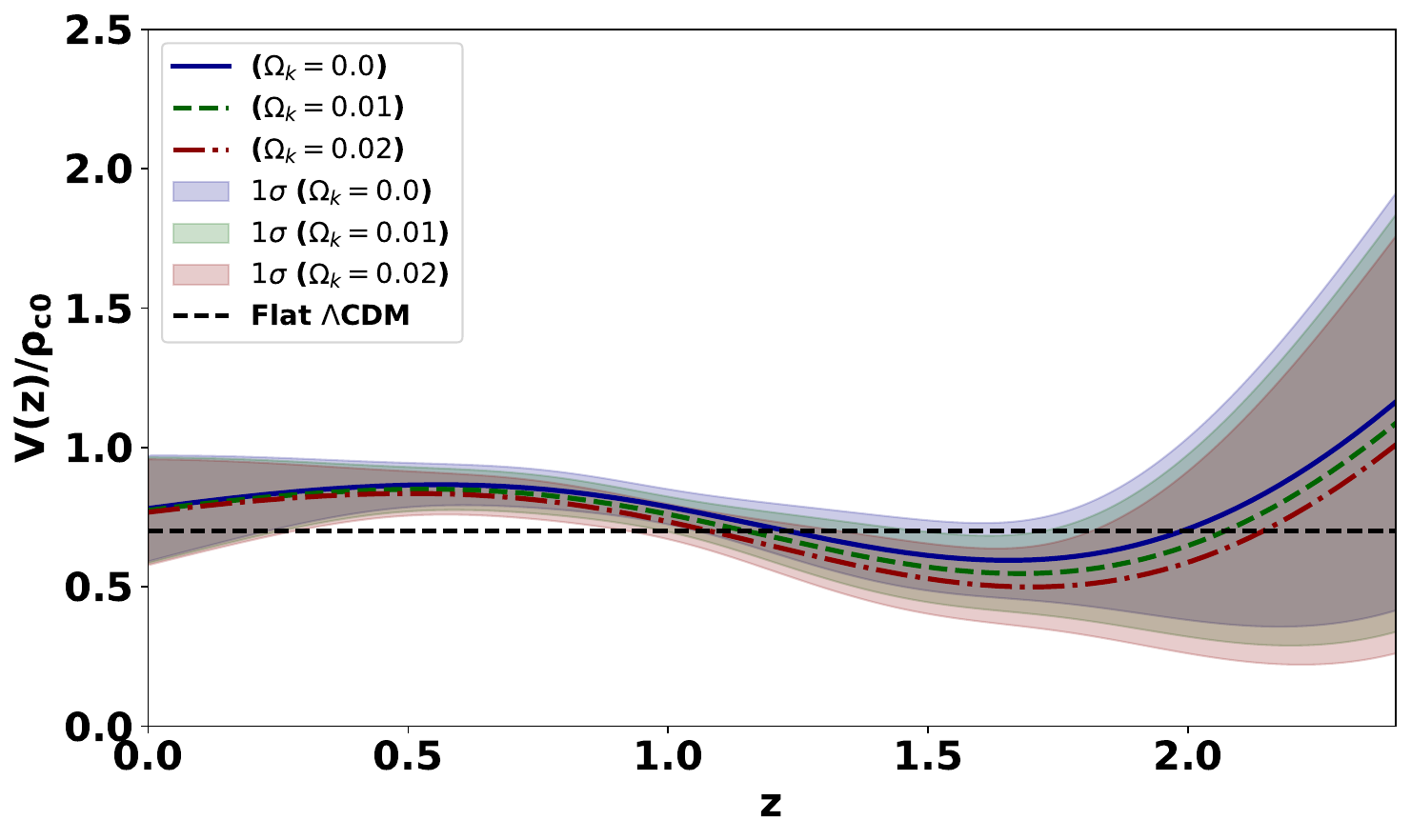}
    
    \medskip
    \includegraphics[width=0.48\textwidth]{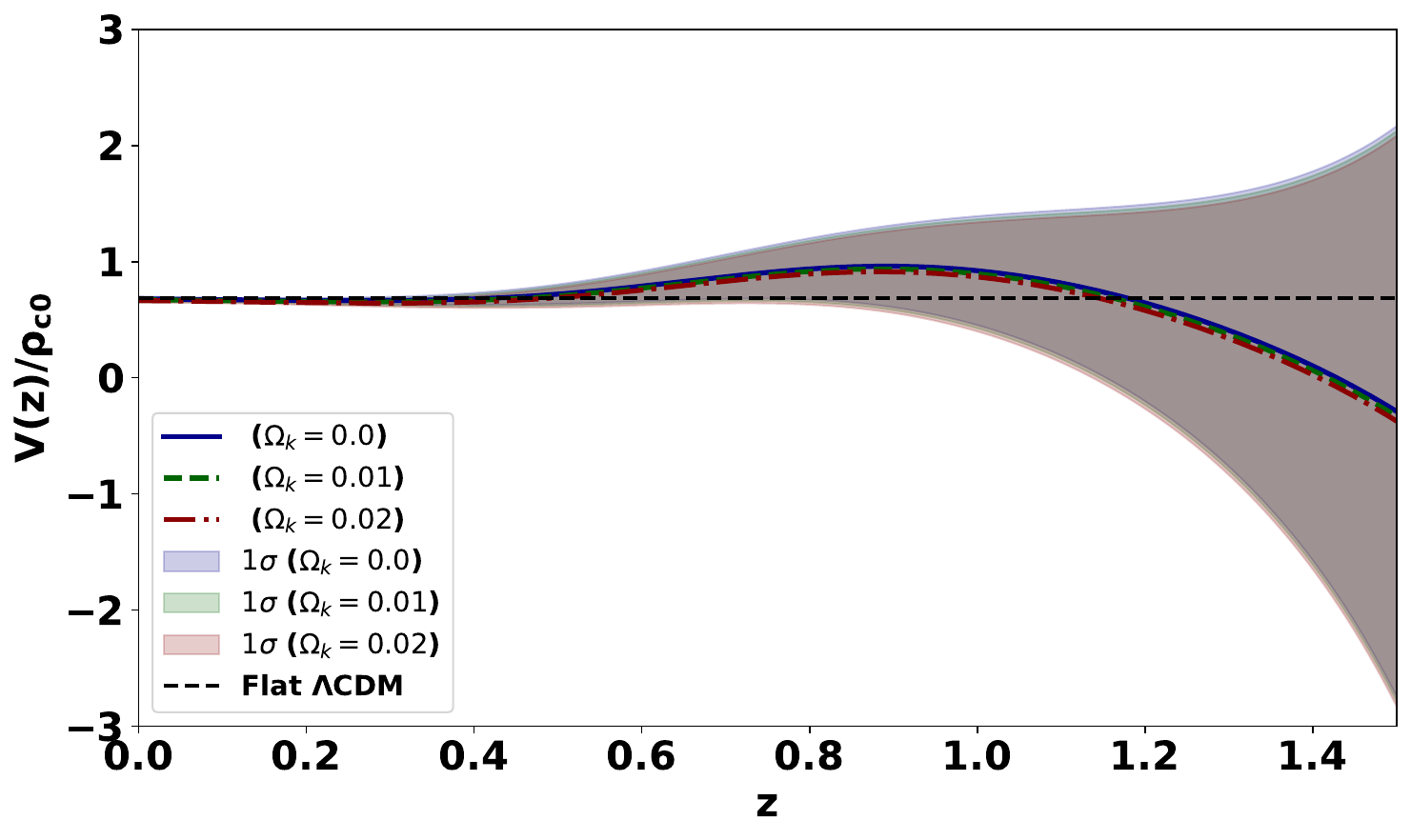}
    \includegraphics[width=0.48\textwidth]{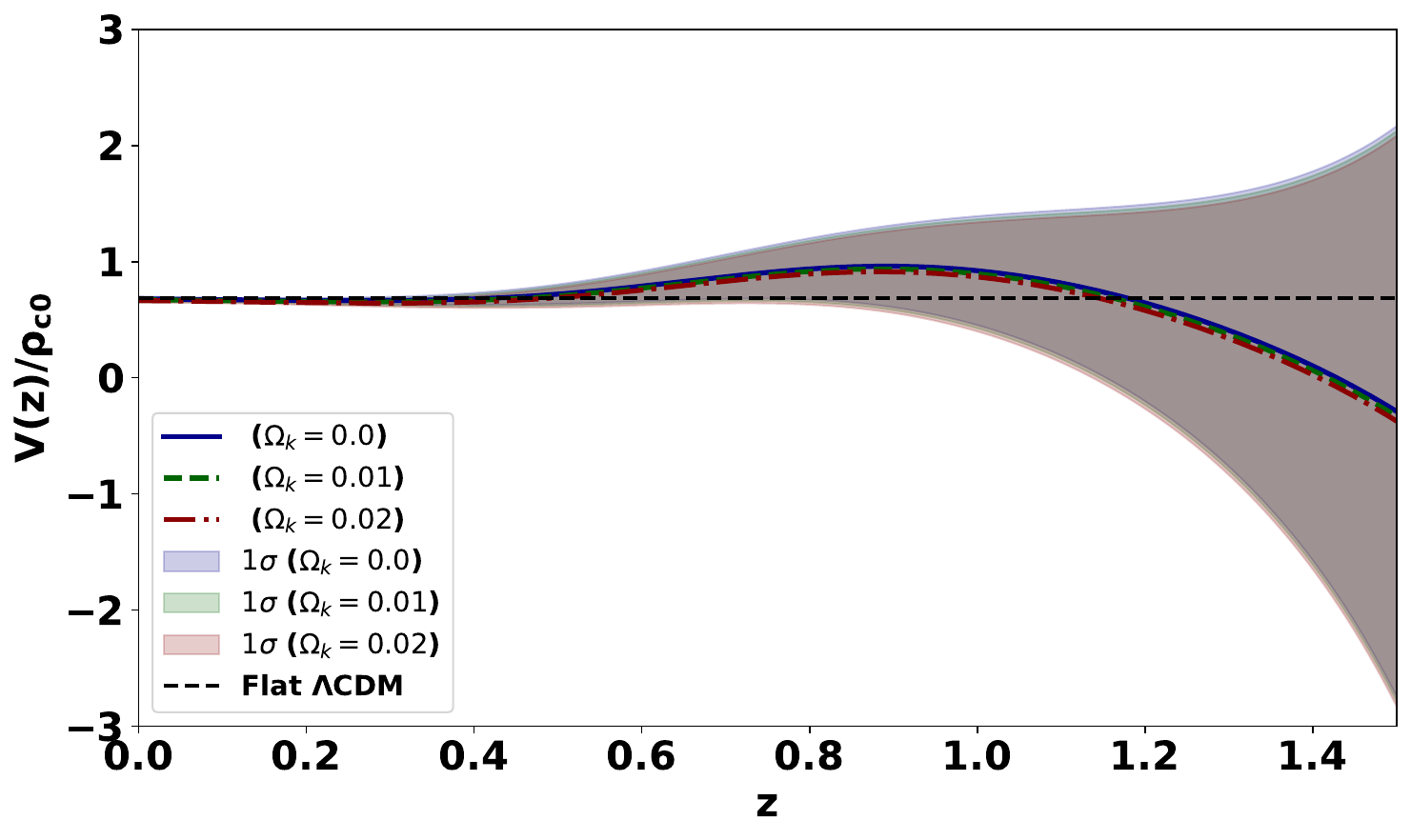}
    \medskip
    
    \includegraphics[width=0.48\textwidth]{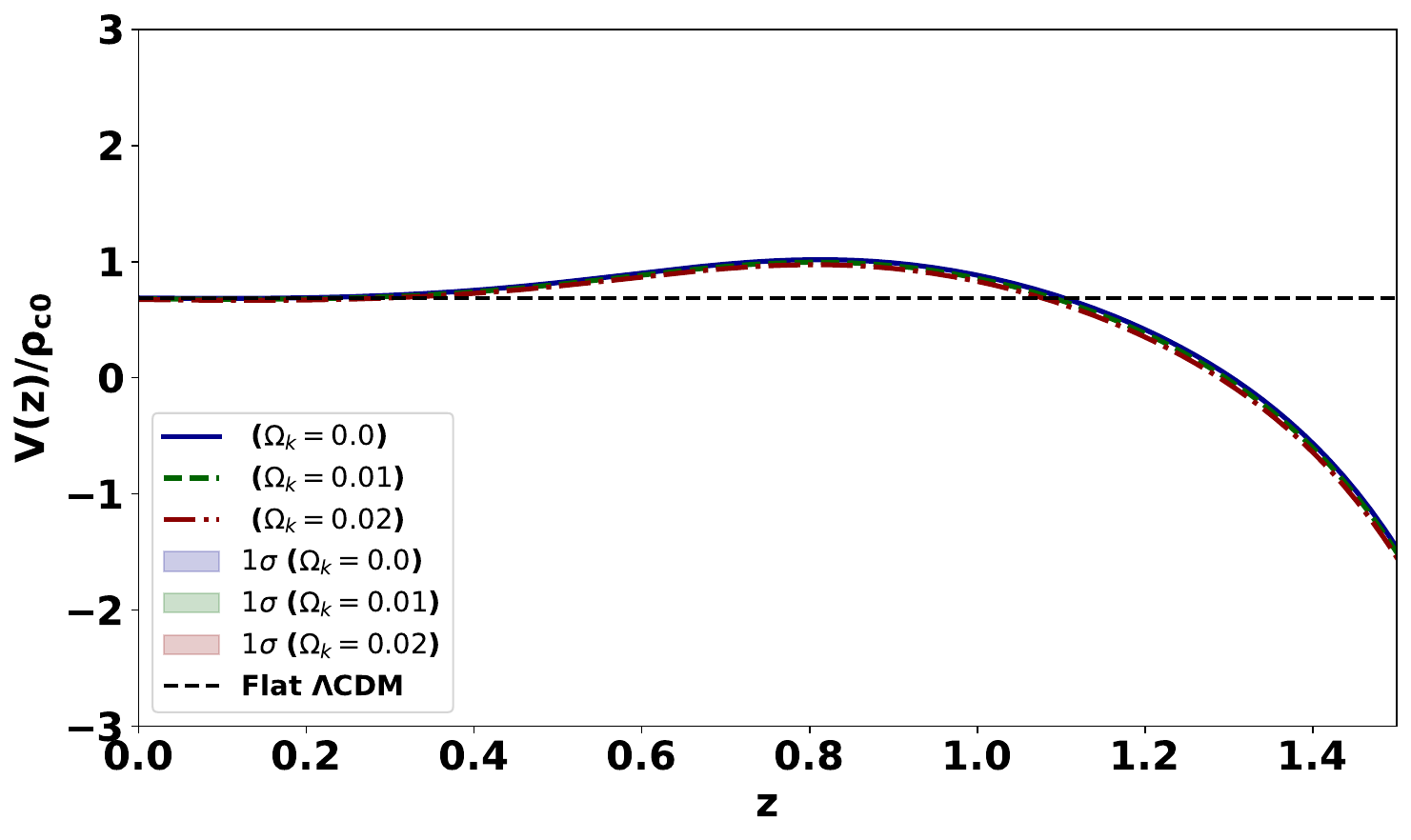}
    \includegraphics[width=0.48\textwidth]{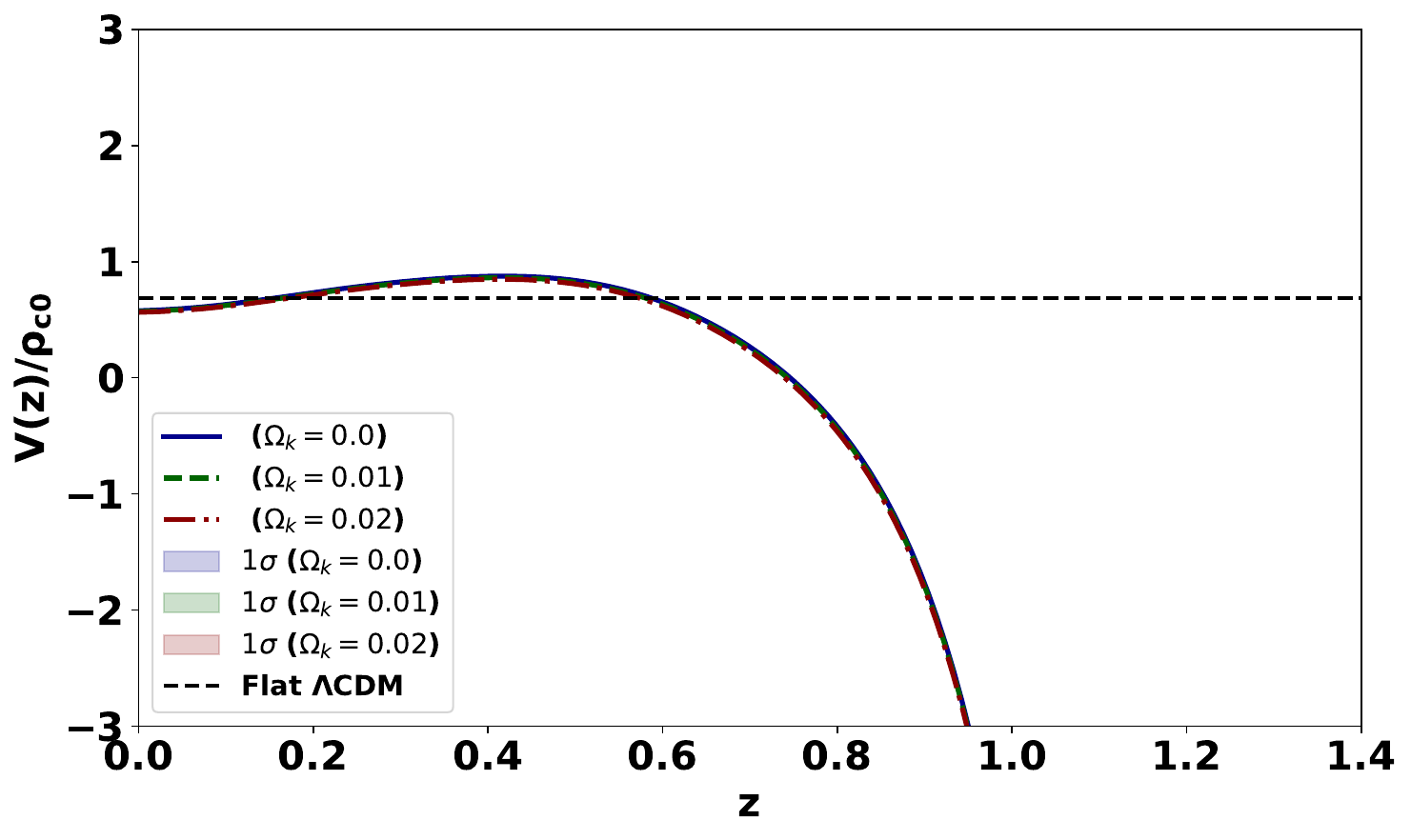}
    
    \caption{Gaussian Process reconstruction of the scalar field potential (~\ref{eq:V}) for different combinations of observational datasets. Top row: CC32 (left) and CC32+DESI DR2 (right). Middle row: CC32+DESI DR2+Pantheon+ (left) and CC32+DESI DR2+Pantheon++CMB (right). Bottom row: CC32+DESI DR2+Union3+CMB (left) and CC32+DESI DR2+DES Y5+CMB (right). Solid curves represent the GP mean reconstruction, while the shaded regions correspond to the $1\sigma$ confidence intervals.}
    \label{fig:V} 
\end{figure*}

Using cosmic chronometer (CC32) data alone \cite{Zhang_2014,Stern:2010cv,Moresco:2012jh,Moresco_2016,10.1093/mnras/stx301,10.1093/mnrasl/slv037}, the reconstruction of the scalar field potential captures the overall trend of the potential but is limited by relatively large uncertainties, particularly at higher redshifts, as shown in the top-left panel of Fig.~\ref{fig:V}. The inclusion of DESI DR2 BAO measurements~\cite{DESI:2025zgx} significantly improves the reconstruction at intermediate redshifts by providing additional direct constraints on the Hubble parameter, as shown in the top-right panel of Fig.~\ref{fig:V}. When Type Ia supernova data from the Pantheon+ compilation~\cite{Scolnic:2017caz,Brout:2022vxf} are incorporated, the reconstruction becomes more stable at low redshift due to their strong sensitivity to the integrated expansion history. The combined datasets thus provide complementary constraints across a wide redshift range, as shown in the middle-left panel of Fig.~\ref{fig:V}.

The reconstructed potential remains nearly constant at low redshift, with $V(z)/\rho_{c,0} \simeq 0.68\text{--}0.70$, consistent with the observed dark energy density in the $\Lambda$CDM model, which is clearly visible in all figures of Fig.~\ref{fig:V}. While mild redshift evolution is allowed within uncertainties, no significant deviation from a constant potential is detected. The uncertainty bands increase at higher redshifts due to the decreasing number and precision of observational data points.

We also examine the impact of including CMB distance priors through compressed shift parameters~\cite{Chen:2018dbv}. Their inclusion does not lead to any noticeable change in the reconstructed potential, indicating that the late time datasets already provide strong constraints on $V(z)$, as seen in the middle-right panel of Fig.~\ref{fig:V}. The consistency of the reconstruction is further confirmed by replacing the Pantheon+ compilation with alternative supernova datasets, namely Union3\cite{Rubin:2023jdq} and DES Y5\cite{DES:2024jxu}, as shown in the bottom-left and bottom-right panels of Fig.~\ref{fig:V}, respectively, while keeping the remaining datasets unchanged. In all cases, the potential remains close to a constant value at late times, with only small variations within the error bands Fig.~\ref{fig:V}. We also find that variations in spatial curvature ($\Omega_k = 0, 0.01, 0.02$) lead to only small shifts within the uncertainty bands.

The reconstructed behavior of $V(z)$ indicates that the dark energy component is dominated by a nearly constant potential, with no statistically significant evidence for evolution within current uncertainties. This is consistent with a slow-roll–like regime in which the scalar field evolves slowly, leading to an equation of state close to $w = -1$, while still allowing for mild dynamical deviations \cite{Wetterich:1987fk,Wetterich:1987fm,Ratra:1987rm,Copeland:2006wr,Bahamonde:2017ize}.

\begin{figure*}[t]
    \centering

    \includegraphics[width=0.49\textwidth]{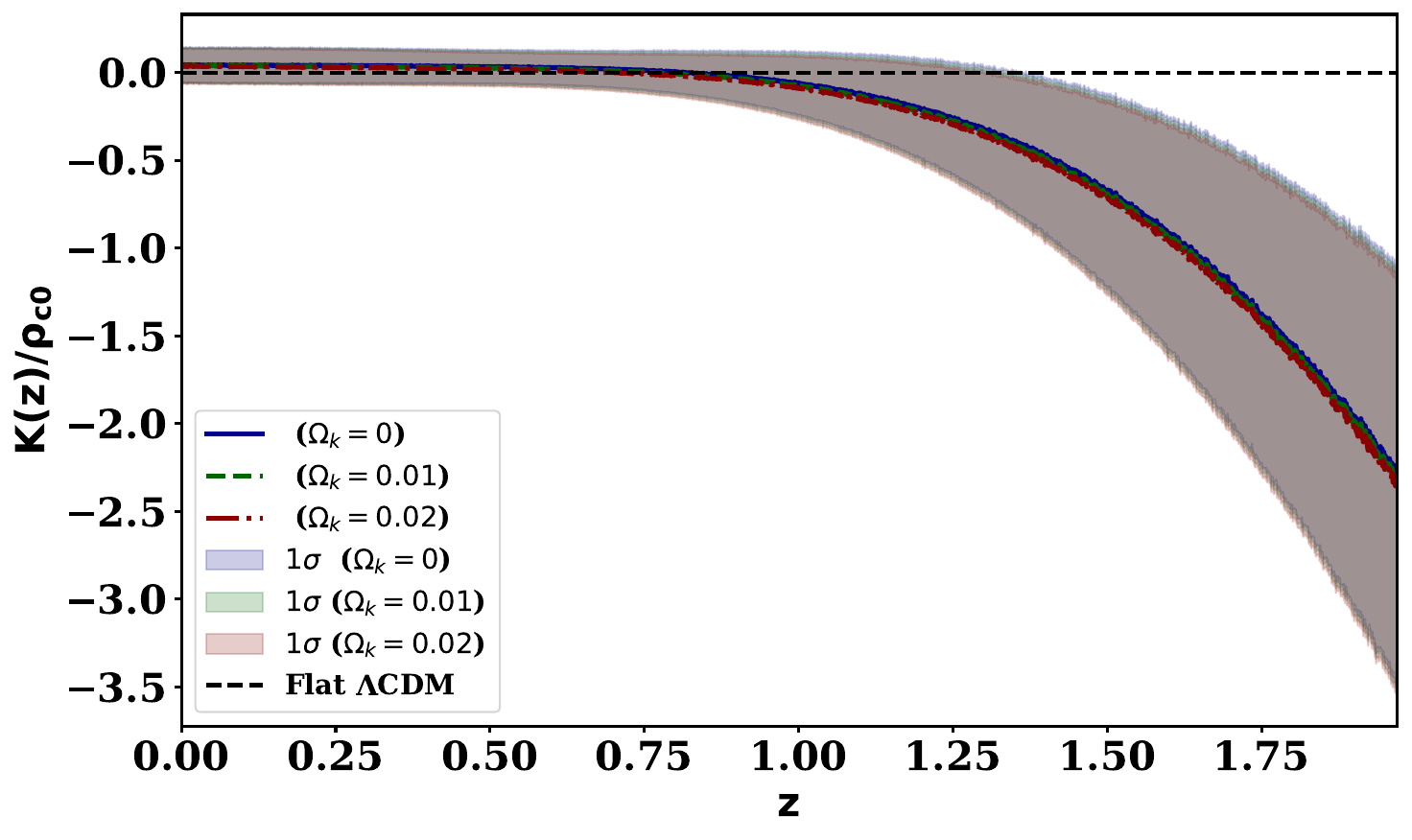}
    \includegraphics[width=0.49\textwidth]{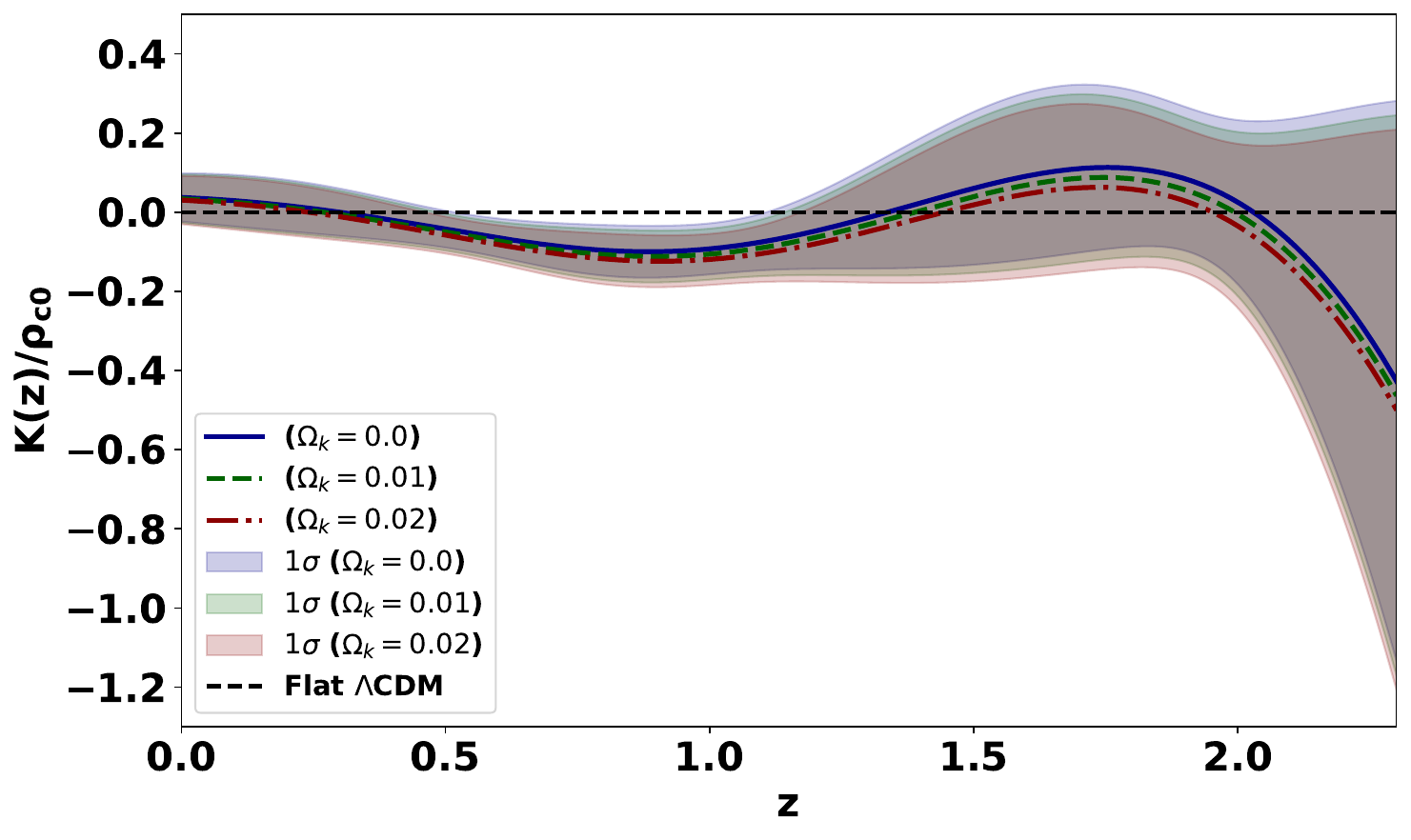}
    
    \medskip
     \includegraphics[width=0.49\textwidth]{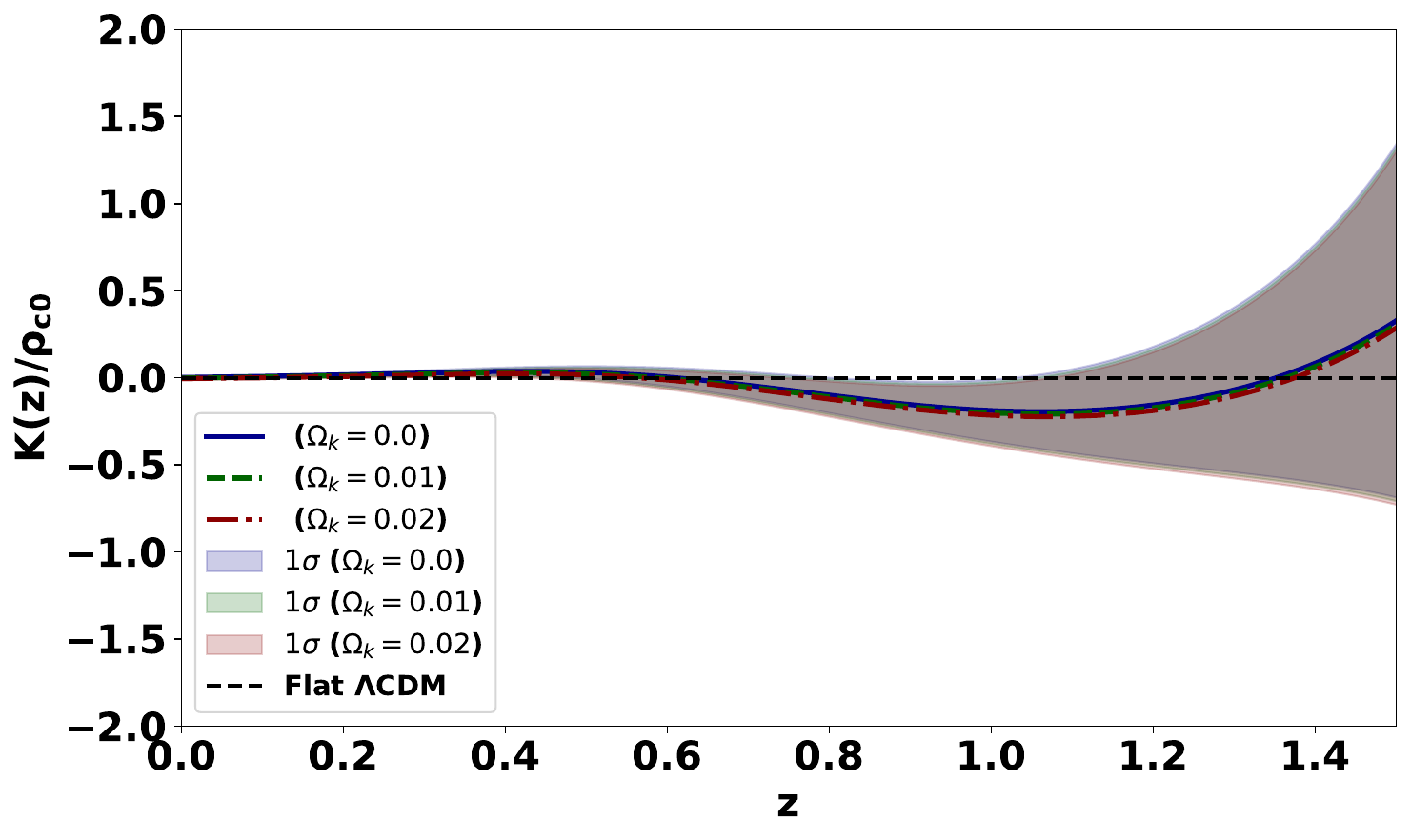}
     \includegraphics[width=0.49\textwidth]{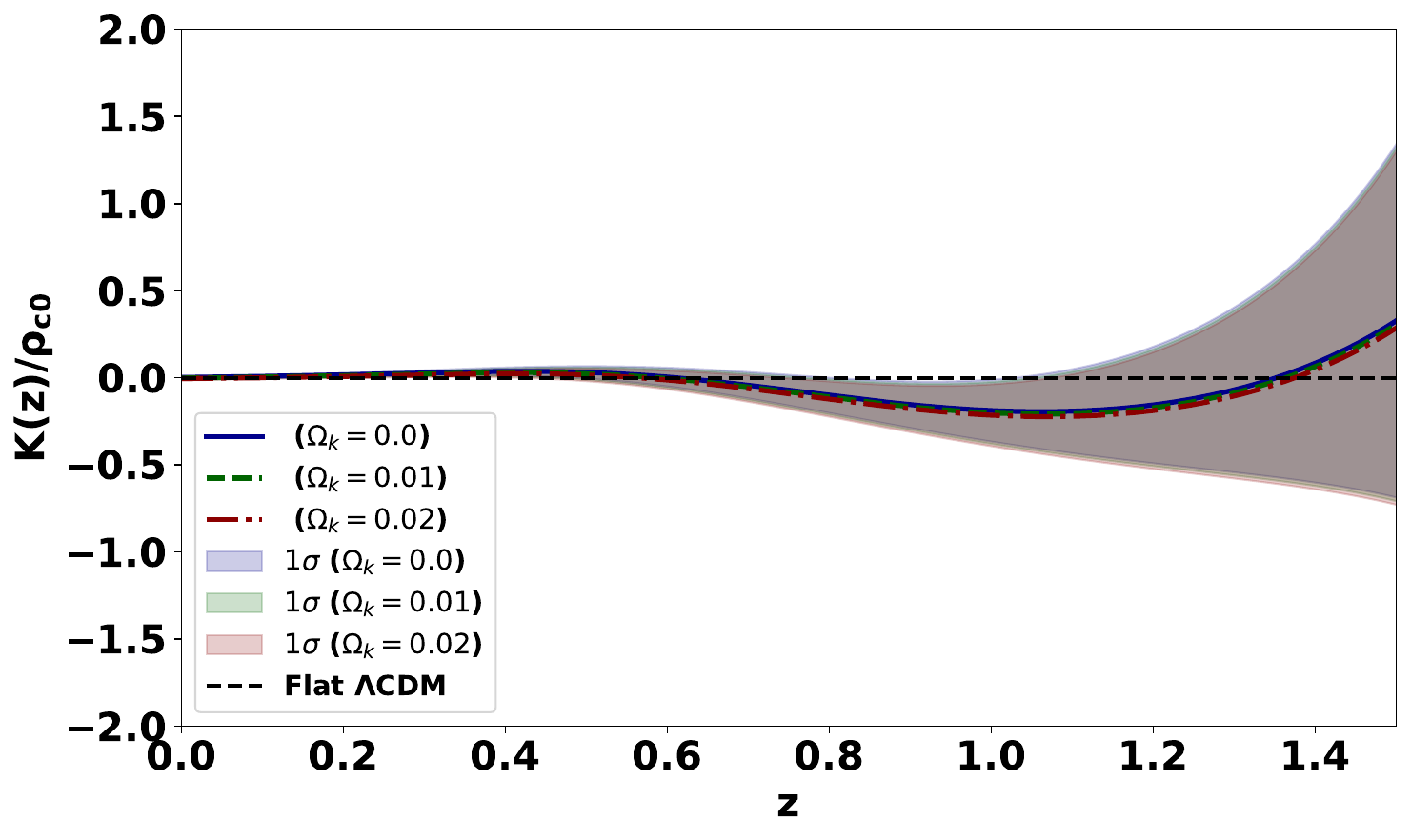}
    \medskip
    
    \includegraphics[width=0.49\textwidth]{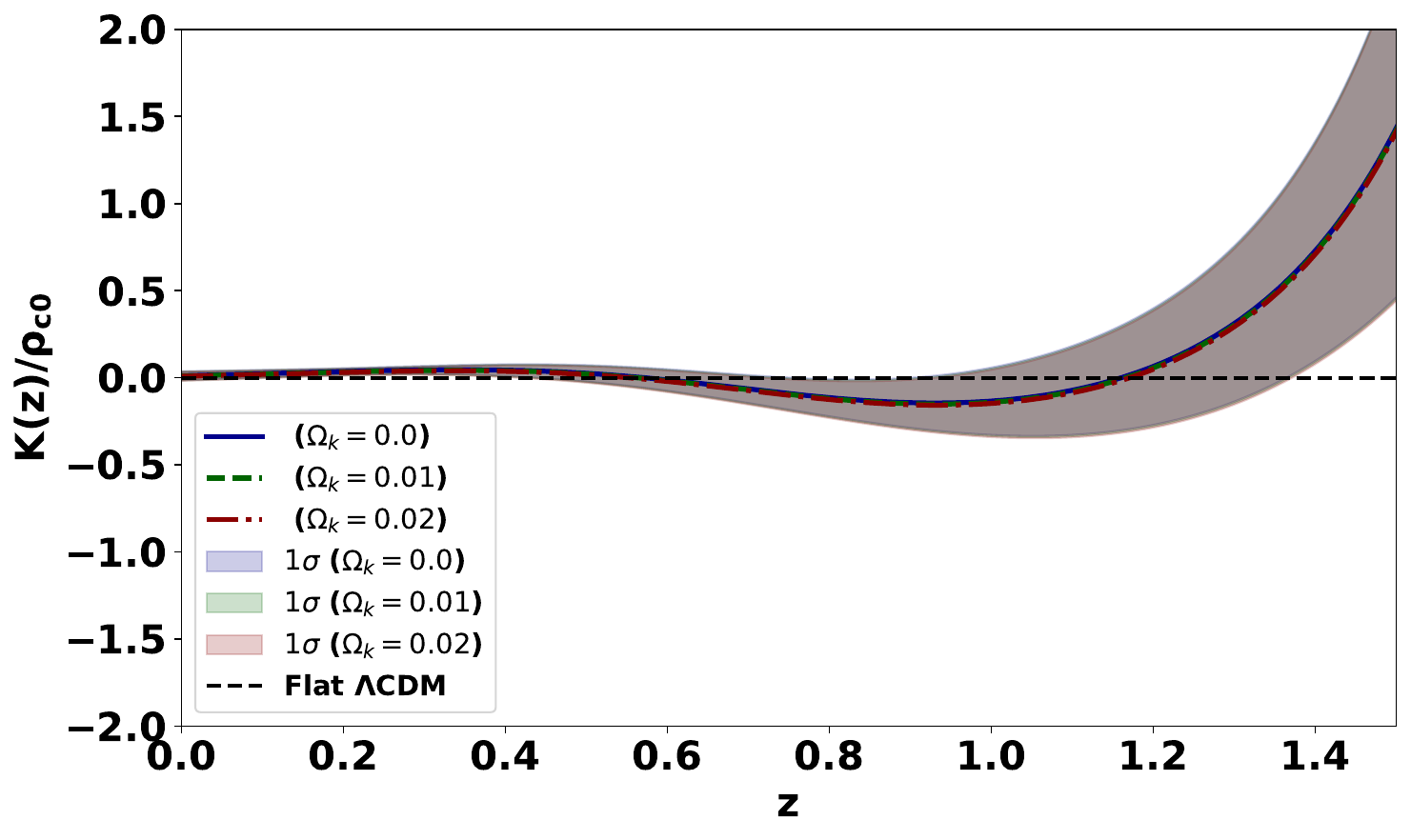}
    \includegraphics[width=0.49\textwidth]{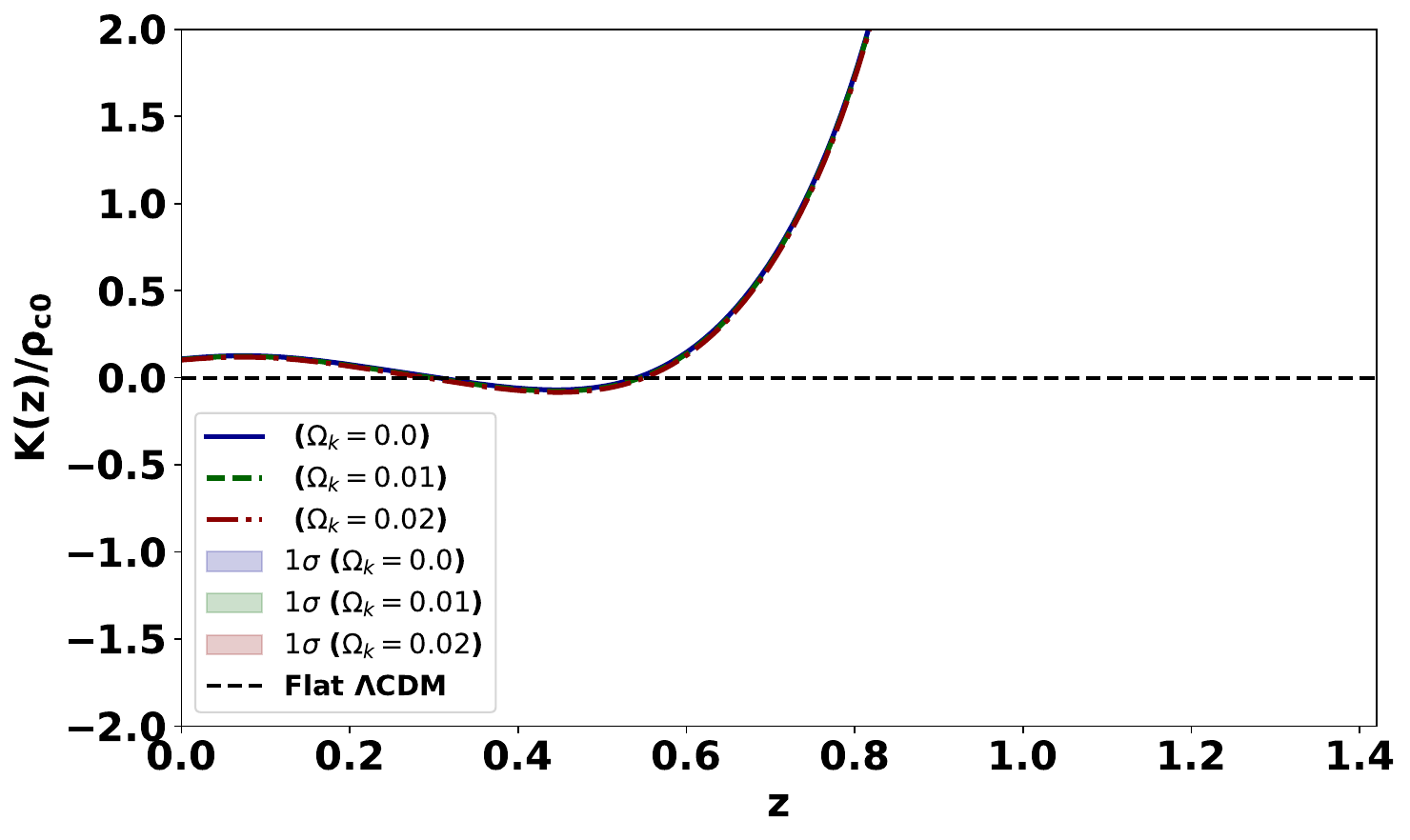}
    
    \caption{Gaussian Process reconstruction of the kinetic term (Eq.~\ref{eq:K}) for different combinations of observational datasets. Top row: CC32 (left) and CC32+DESI DR2 (right). Middle row: CC32+DESI DR2+Pantheon+ (left) and CC32+DESI DR2+Pantheon++CMB (right). Bottom row: CC32+DESI DR2+CMB combined with Union3 (left) and DES Y5 (right) supernova datasets. Solid curves represent the GP mean reconstruction, while the shaded regions correspond to the $1\sigma$ confidence intervals.}
    \label{fig:K}
\end{figure*}

\subsection{Reconstruction of the Kinetic Term of scalar field}
\label{sec:K}
Having reconstructed the potential component of the scalar field in the previous section, we now turn to the reconstruction of the kinetic term (\ref{eq:K}). Together, the kinetic and potential contributions determine the dynamical behavior of the scalar field responsible for the late-time acceleration of the Universe. Using cosmic chronometer (CC32) \cite{Zhang_2014,Stern:2010cv,Moresco:2012jh,Moresco_2016,10.1093/mnras/stx301,10.1093/mnrasl/slv037} data alone, the reconstructed kinetic term remains close to zero but exhibits relatively large uncertainties, particularly at higher redshifts where the data become sparse, as shown in the top-left panel of Fig.~\ref{fig:K}. The reconstruction also shows a tendency to cross below the $K(z)=0$ line at smaller redshifts, although this behavior is not statistically significant given the broad uncertainty bands. The inclusion of DESI DR2 BAO ~\cite{DESI:2025zgx} measurements significantly improves the stability of the reconstruction at intermediate redshifts by providing additional constraints on $H(z)$ and its derivatives, as shown in the top-right panel of Fig.~\ref{fig:K}. In particular, the reconstructed kinetic term exhibits a mild crossing below the $K(z)=0$ line over the approximate redshift range $0.5 \lesssim z \lesssim 1.3$, indicating a feature primarily driven by the DESI DR2 data \cite{DESI:2025zgx,DESI:2025wyn,DESI:2025fii}. 

Incorporating Type Ia supernova data from the Pantheon+ compilation further reduces uncertainties at low redshift, leading to a smoother and more tightly constrained reconstruction below $z \lesssim 1$, while preserving the DESI-induced feature, as shown in the middle-left panel of Fig.~\ref{fig:K}. The reconstructed kinetic term remains subdominant compared to the potential contribution, with $K(z)$ staying close to zero within uncertainties throughout the observed redshift range. Although mild fluctuations appear at higher redshifts due to amplified derivative uncertainties and sparse data coverage, no statistically significant deviation from a small kinetic contribution is detected. We also assess the impact of including CMB distance priors ~\cite{Chen:2018dbv} through compressed shift parameters. Similar to the case of the potential reconstruction, their inclusion does not produce any noticeable visible change in the reconstructed kinetic term, indicating that the late time datasets already provide strong constraints on the expansion history and its derivatives, as shown in the middle-right panel of Fig.~\ref{fig:K}. 

The robustness of the reconstruction is further confirmed by replacing the Pantheon+ compilation with the Union3\cite{Rubin:2023jdq} and DES Y5 \cite{DES:2024jxu}supernova datasets. In both cases, the kinetic term remains close to zero, while the DESI-induced crossing near $z \sim 0.5$ persists, as shown in the bottom-left and bottom-right panels of Fig.~\ref{fig:K}, respectively. We also find that variations in spatial curvature ($\Omega_k = 0, 0.01, 0.02$) introduce only small shifts within the uncertainty bands.
\subsection{Reconstruction of $\lambda$}
\label{sec:lambda}
We reconstruct the slope parameter of the effective scalar field potential, defined in Eq.~(\ref{eq:lambda}), directly from observational data in a model-independent framework. GP is used to reconstruct the Hubble parameter $H(z)$ together with its derivatives, from which the scalar field kinetic and potential energy densities, $K(z)$ and $V(z)$, are obtained. The scalar field velocity is then determined through $\dot{\phi}^{,2}=2K(z)$, enabling the reconstruction of the potential slope parameter $|\lambda(z)|$ without assuming any specific functional form of the scalar field potential. Since $|\lambda(z)|$ is a nonlinear function of the jointly reconstructed
$H(z)$ and its derivatives, involving a ratio whose denominator may become small, its probability distribution is generally non-Gaussian. Consequently, conventional first-order propagation of GP uncertainties can yield symmetric confidence intervals extending below zero, despite $|\lambda(z)|$ being non-negative by construction. We therefore propagate
the full joint GP posterior using Monte Carlo realizations and evaluate $|\lambda(z)|$ for each realization. The reconstructed distribution is then
summarized by its median together with the 16th and 84th quantiles of the Monte Carlo sample, yielding an asymmetric confidence interval that naturally respects the non-negative nature of $|\lambda(z)|$ and provides a
statistically consistent representation of the propagated uncertainties (\ref{eq:lam_med1},\ref{eq:lam_med2}).
\begin{figure*}[t]
    \centering

    \includegraphics[width=0.32\textwidth]{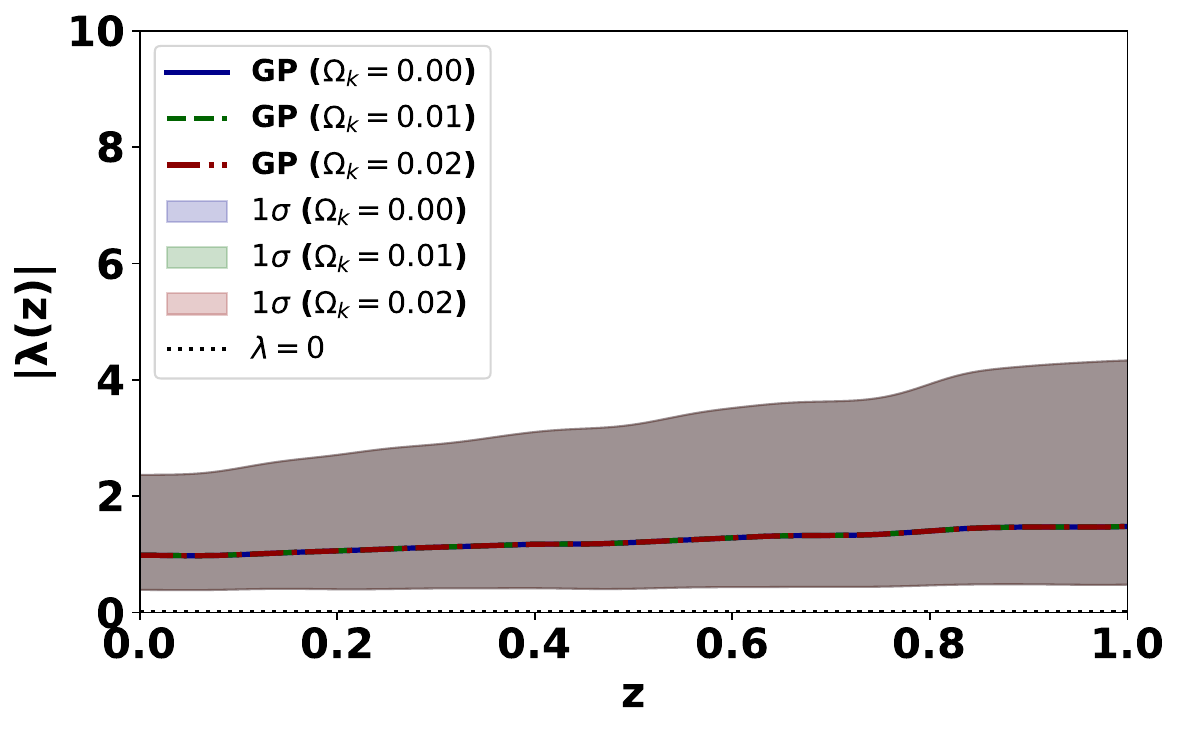}
    \hfill
    \includegraphics[width=0.32\textwidth]{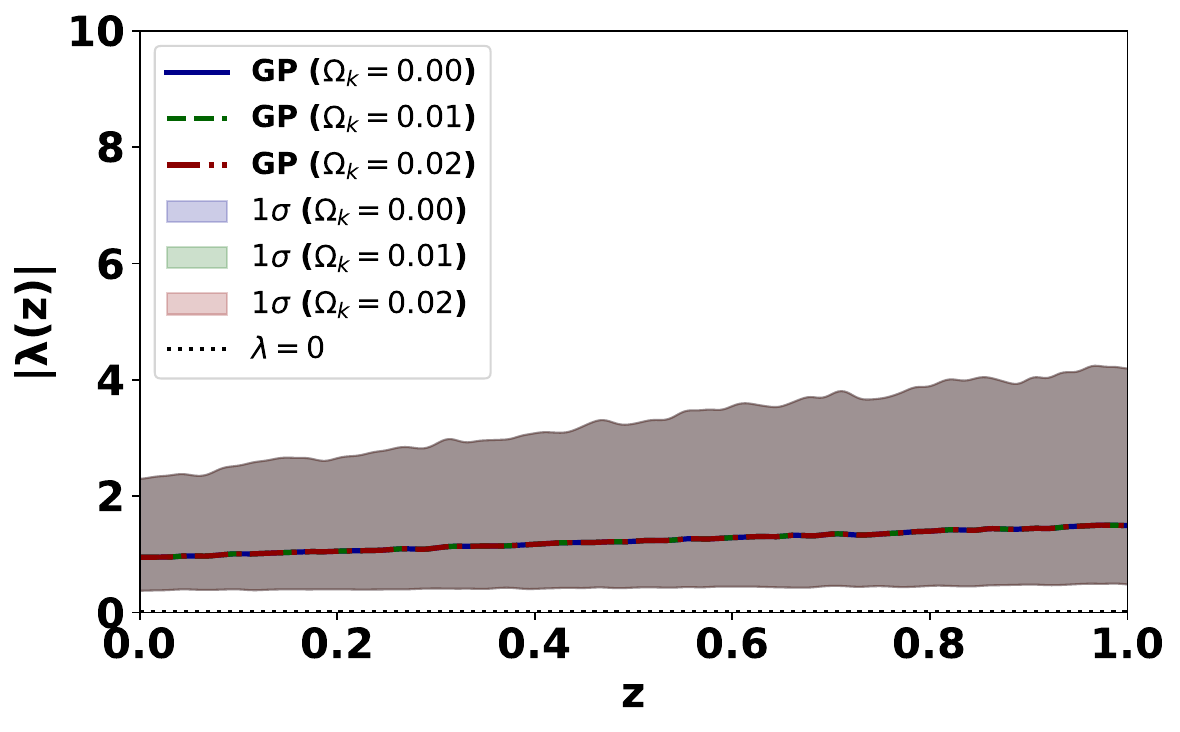}
    \hfill
    \includegraphics[width=0.32\textwidth]{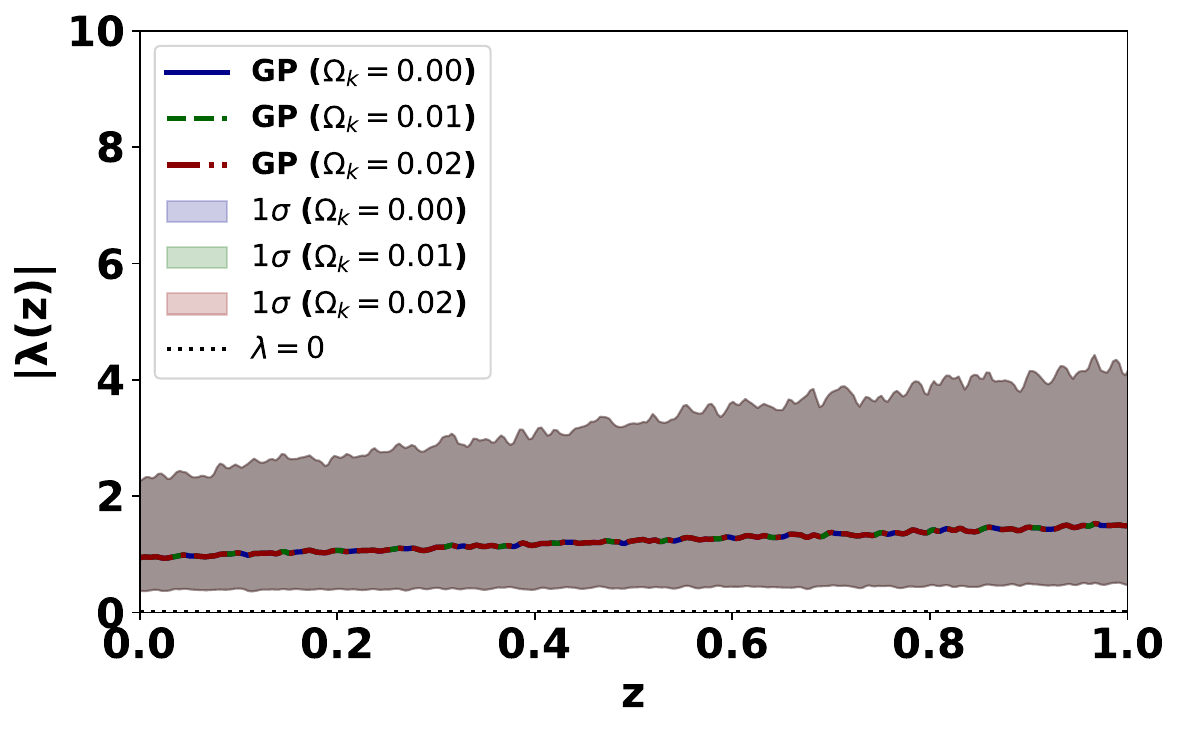}

    \caption{Gaussian Process reconstruction of the slope parameter $|\lambda(z)|$ [Eq.~(\ref{eq:lambda})] using different combinations of observational datasets. The left, middle, and right panels correspond to the CC32 + DESI DR2 + Pantheon+ + CMB, CC32 + DESI DR2 + Union3 + CMB, and CC32 +DESI DR2+ DES Y5 + CMB datasets, respectively. The solid curves represent the median Monte Carlo reconstruction, while the shaded regions denote the 16th--84th percentile (68\%) confidence intervals.}
    \label{fig:lambda}
\end{figure*}

Figure~\ref{fig:lambda} presents the reconstructed evolution of the slope parameter $|\lambda(z)|$ for the three supernova compilations. In all cases, the median reconstruction exhibits only mild evolution with redshift and remains remarkably insensitive to the assumed spatial curvature, with the results for $\Omega_k=0.00$, $0.01$, and $0.02$ nearly overlapping throughout the reconstructed range. We quote the reconstructed present day values of $|\lambda(z)|$ for the spatially flat case ($\Omega_k=0$), since the results exhibit only a weak dependence on the assumed spatial curvature and remain nearly unchanged for $\Omega_k=0.01$ and $\Omega_k=0.02$. The reconstructed present-day median values of $|\lambda(z)|$ are $0.971^{+1.193}_{-0.505}$, $1.007^{+1.389}_{-0.625}$, and $0.929^{+1.402}_{-0.661}$ for the CC32+DESI DR2+CMB combinations with Pantheon+, Union3, and DES Y5, respectively (Fig\ref{fig:lambda}).
Although the overall evolution is qualitatively similar for all three datasets, the uncertainty increases rapidly toward higher redshift because $|\lambda(z)|$ depends on $dV/dz$, which in turn requires the second derivative of the Hubble parameter, $H''(z)$. Since $H(z)$ is reconstructed from luminosity-distance measurements, this is effectively equivalent to estimating the third derivative of the distance--redshift relation, $D'''(z)$. Consequently, the uncertainties are substantially larger than those of lower-order quantities such as $H(z)$, $w(z)$, $K(z)$, and $V(z)$, so current observations do not yet tightly constrain the magnitude of the potential slope over the full redshift range.

In contrast, the reconstruction of the curvature parameter $\Gamma(z)$ [Eq.~(\ref{eq:Gamma})] is considerably more challenging, as it involves higher-order derivatives of the reconstructed expansion history than $|\lambda(z)|$, effectively requiring derivatives up to $H'''(z)$, corresponding to the fourth derivative of the distance--redshift relation, $D''''(z)$. Although the mean reconstruction exhibits a smooth evolution with mild oscillatory behavior around values of order unity, the propagated uncertainties are substantially amplified, resulting in very broad confidence intervals. Consequently, the present data do not provide statistically meaningful constraints on $\Gamma(z)$ or allow a distinction between different classes of scalar field potentials, including scaling~\cite{Copeland:1997et}, tracker~\cite{Zlatev:1998tr,Steinhardt:1999nw}, and thawing~\cite{Scherrer:2007pu} models. For this reason, we do not present the reconstruction of $\Gamma(z)$ in this work.
\begin{figure*}[t]
    \centering

    \includegraphics[width=0.49\textwidth]{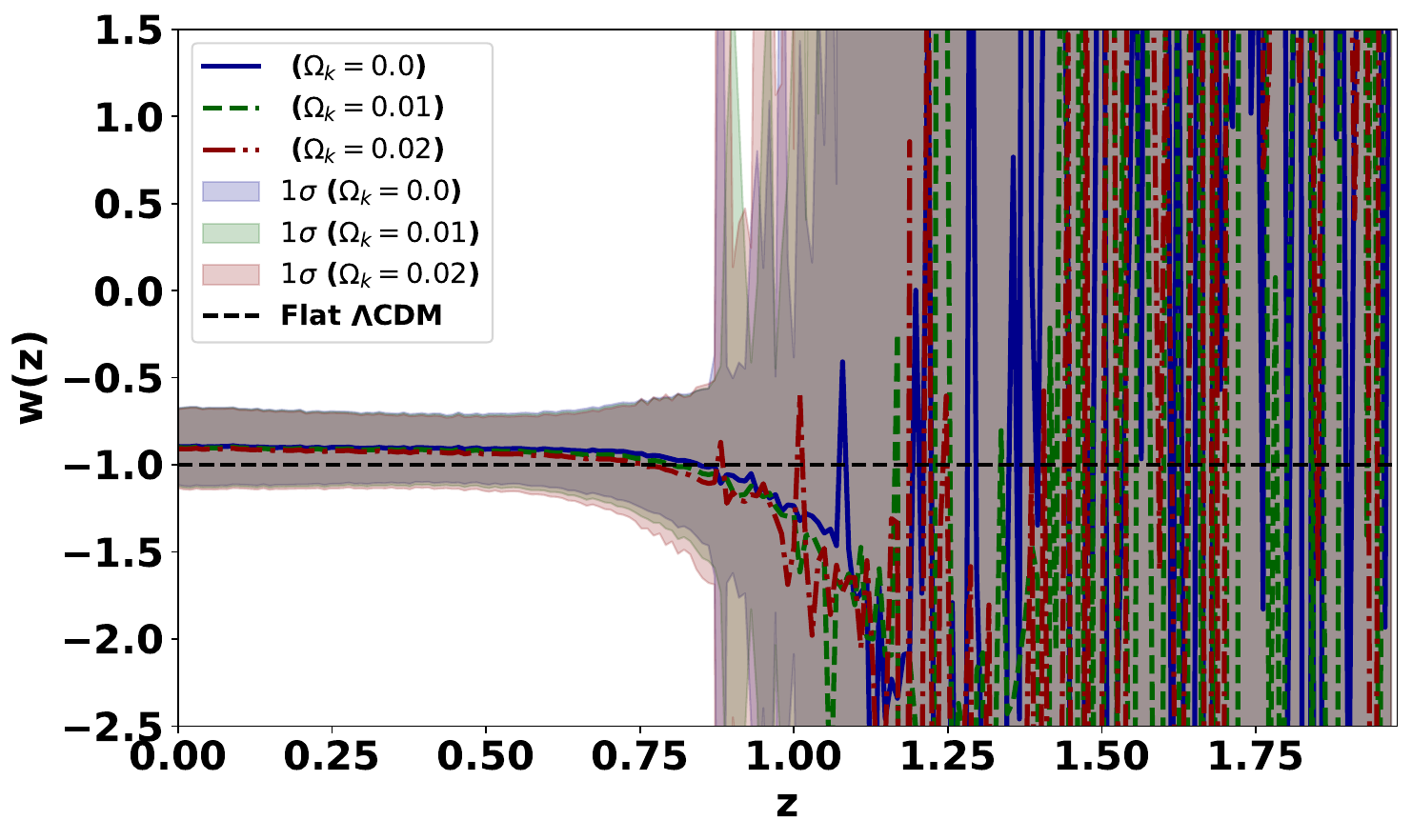}
    \includegraphics[width=0.49\textwidth]{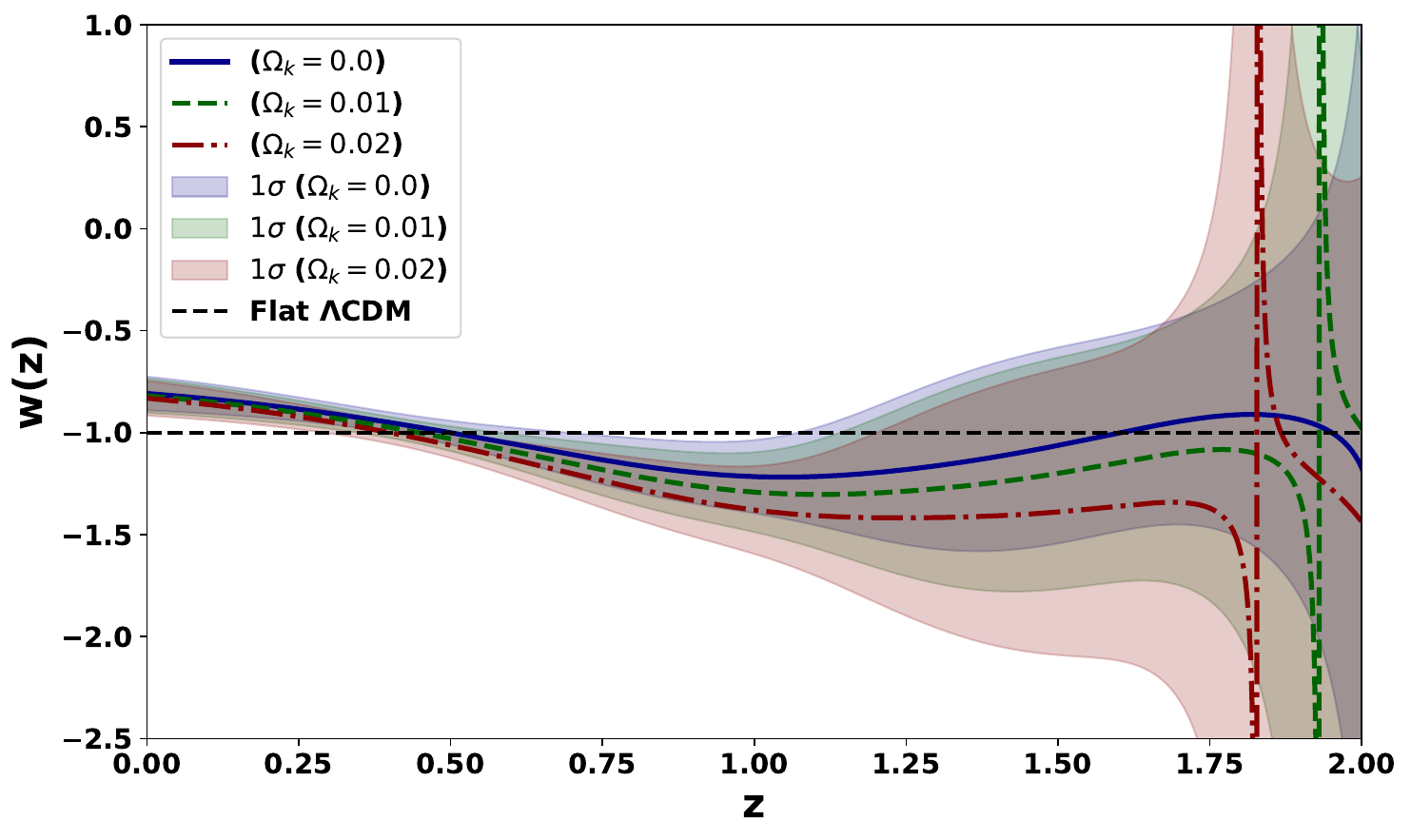}
    \medskip
    \includegraphics[width=0.49\textwidth]{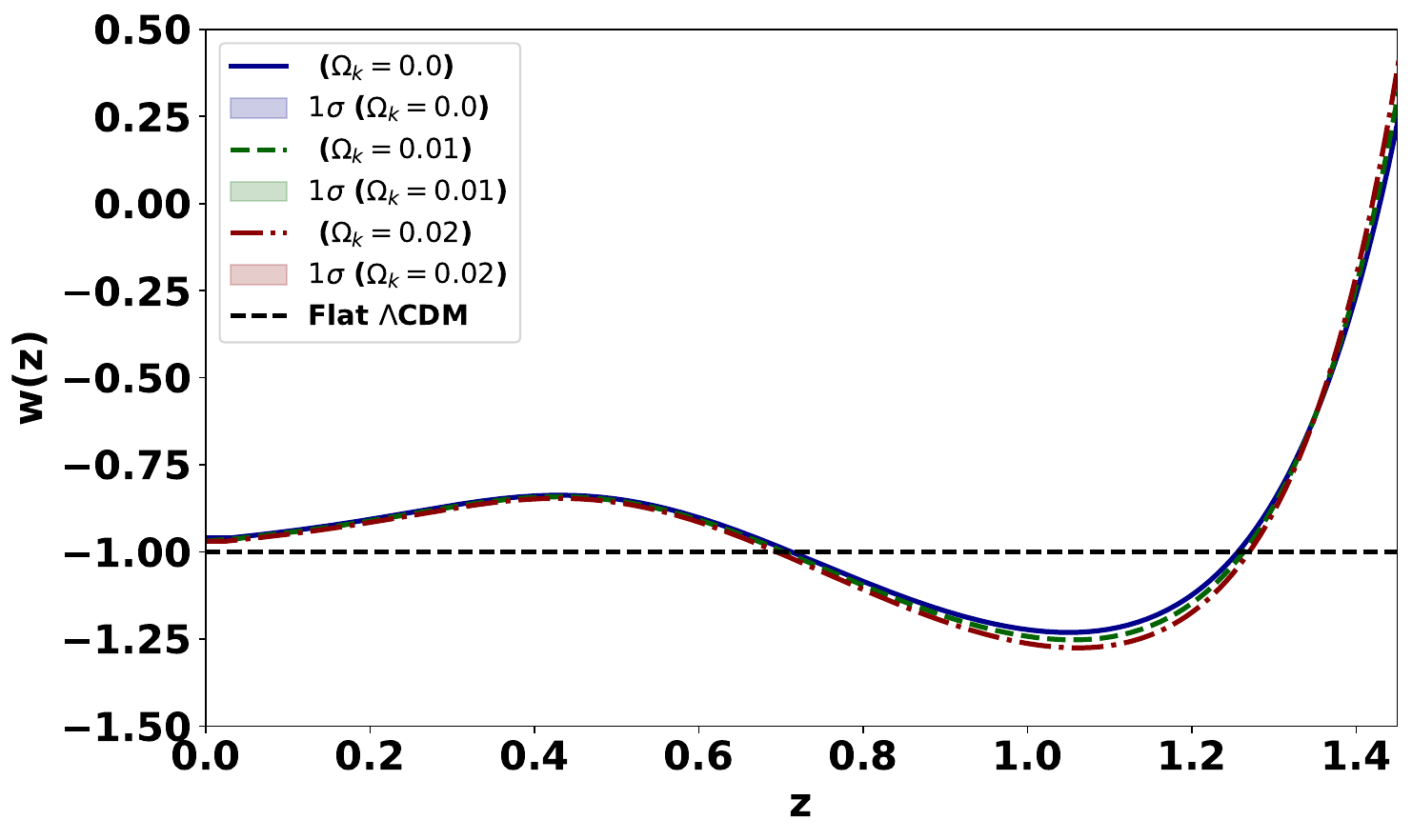}
    \includegraphics[width=0.49\textwidth]{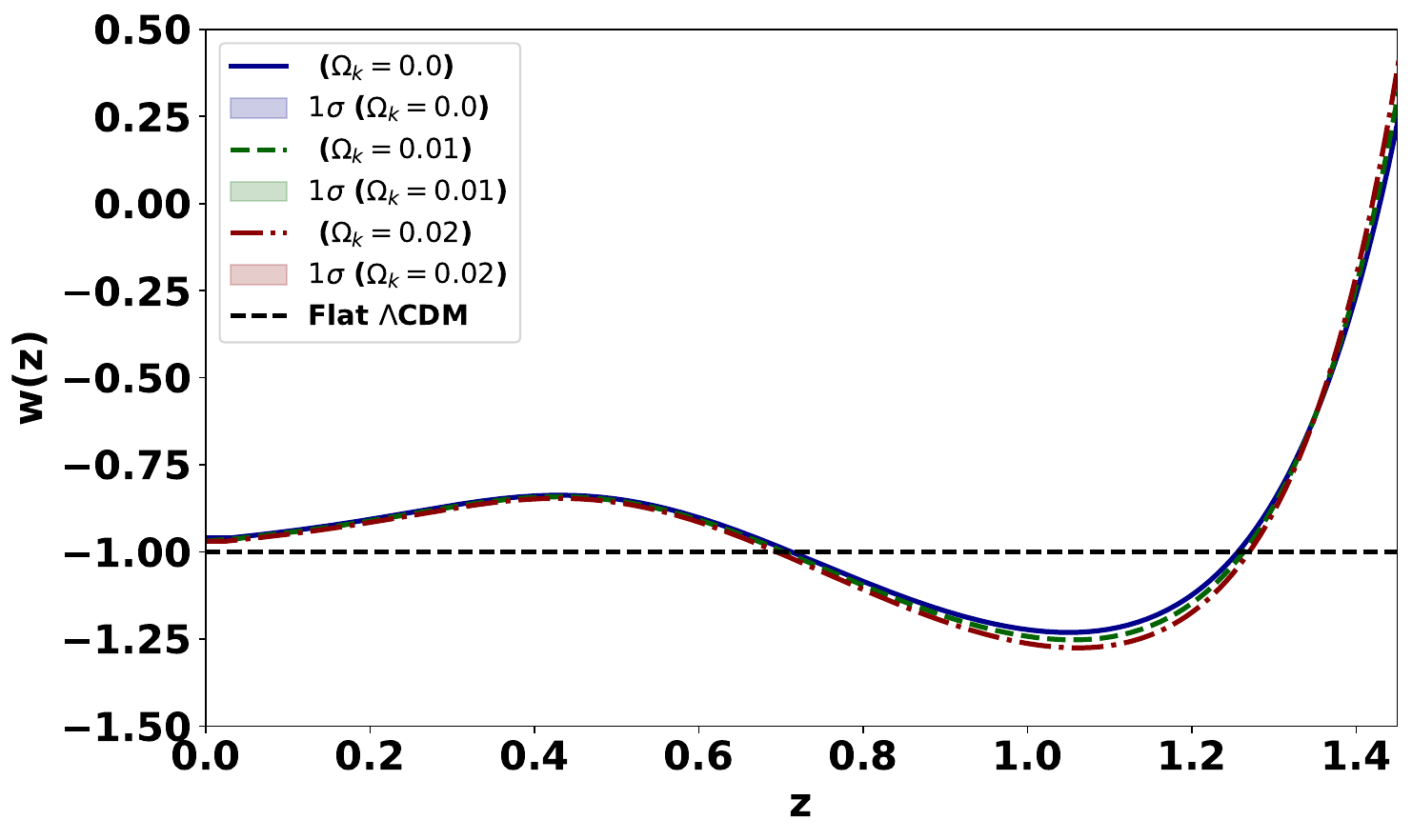}
    \medskip
    \includegraphics[width=0.49\textwidth]{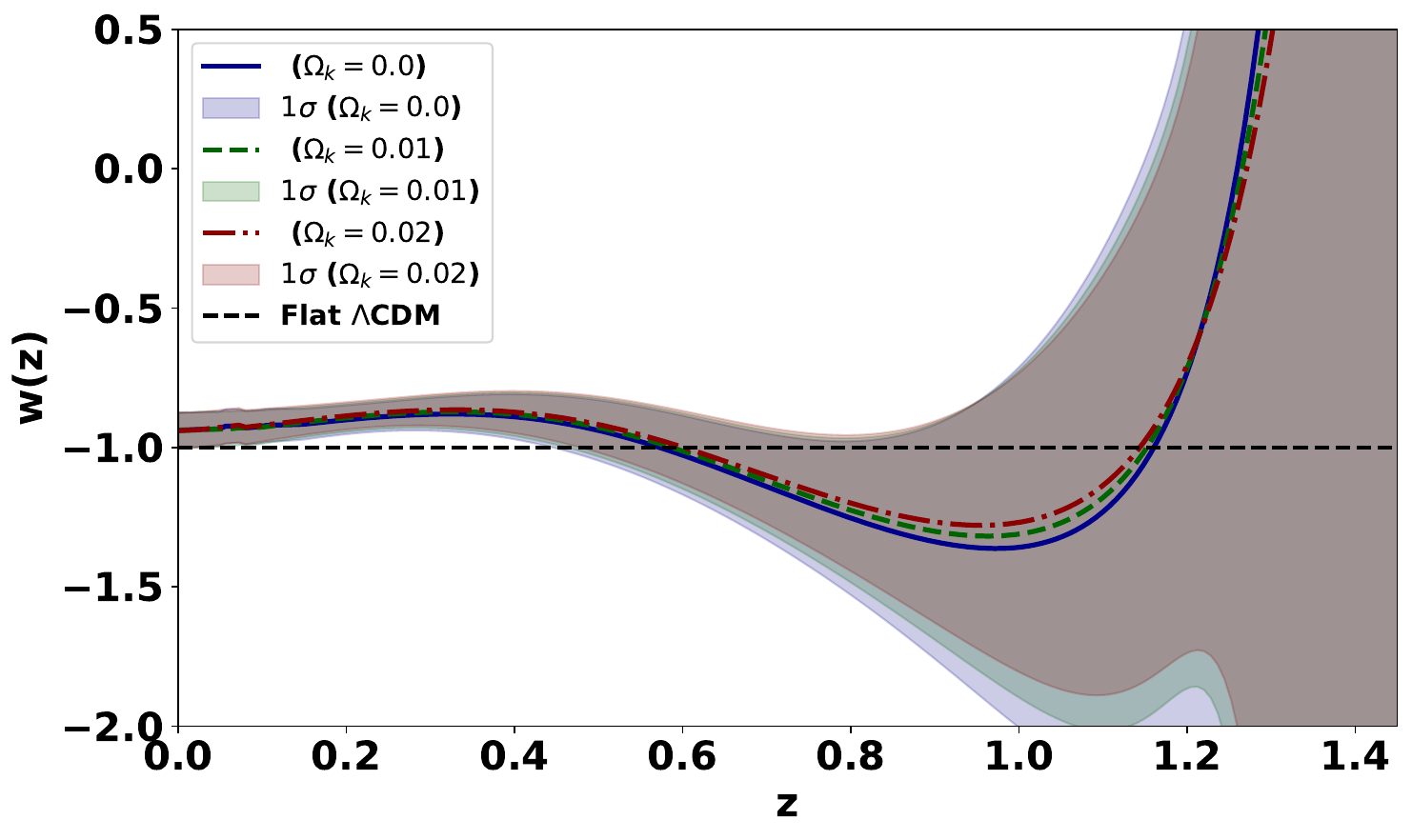}
    \includegraphics[width=0.49\textwidth]{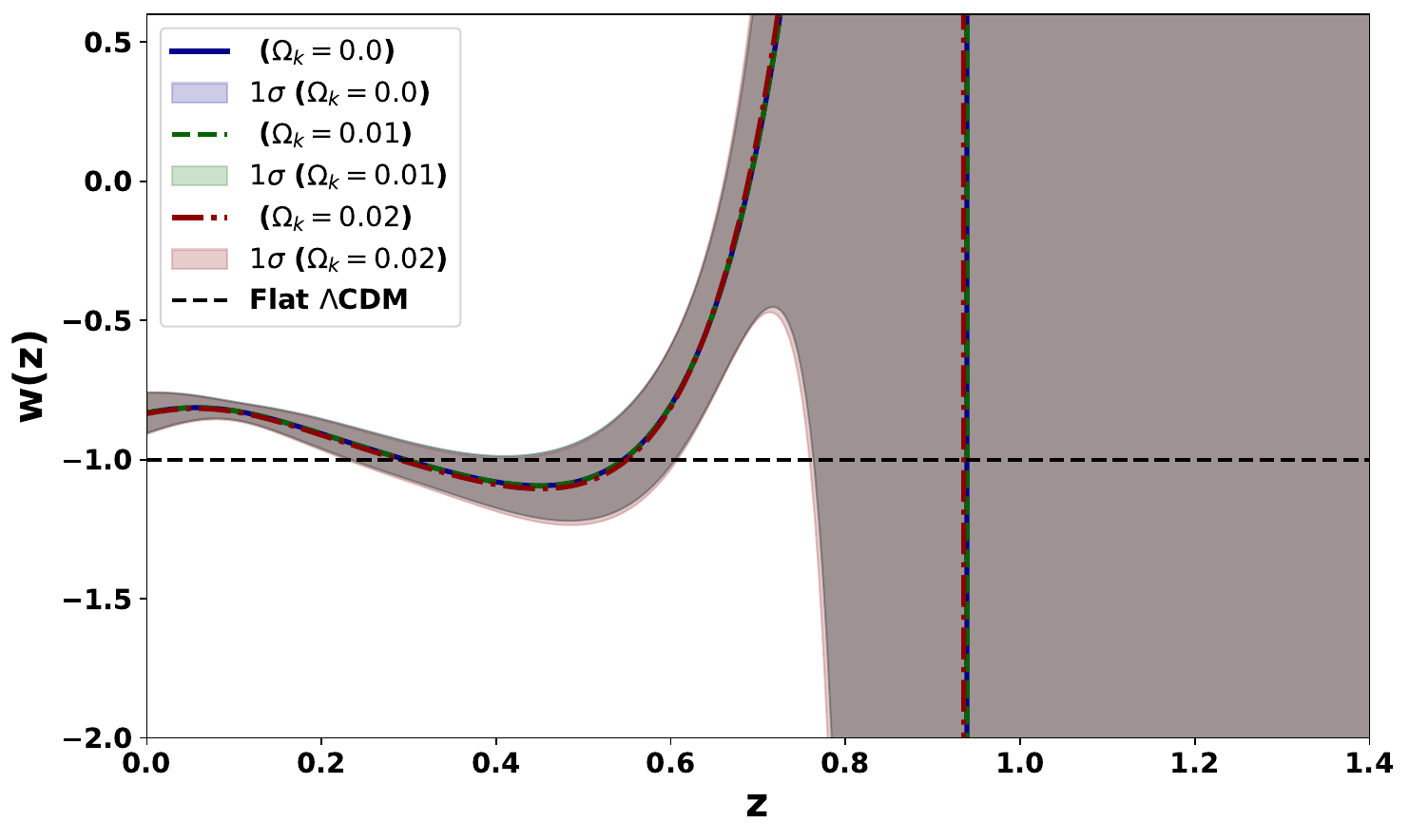}

    \caption{Gaussian Process reconstruction of the equation of state $w(z)$ (~\ref{eq:w}) for progressively combined datasets. Top row: CC32 (left) and CC32+DESI DR2 (right). Middle row: CC32+DESI DR2+Pantheon+ (left) and CC32+DESI DR2+Pantheon++CMB (right). Bottom row: CC32+DESI DR2+CMB combined with Union3 (left) and DES Y5 (right) supernova datasets. Solid curves represent the GP mean reconstruction, while the shaded regions correspond to the $1\sigma$ confidence intervals.}
    \label{fig:w}
\end{figure*}

\subsection{Reconstruction of the Equation of State $w(z)$}
\label{sec:w}
The equation of state parameter $w(z)$ provides the most direct probe of the dynamical nature of dark energy. We are strictly following the definition of the equation of state $w$ defined by Eq.~(\ref{eq:w}). Within the scalar field framework, it can be expressed in terms of the kinetic and potential energy contributions (\ref{eq:w}), thereby directly linking the reconstructed expansion history to the underlying field dynamics. A value $w = -1$ corresponds to the cosmological constant limit, while any deviation from this value signals the presence of dynamical dark energy. 

We begin with the reconstruction based on the CC-32 dataset\cite{Zhang_2014,Stern:2010cv,Moresco:2012jh,Moresco_2016,10.1093/mnras/stx301,10.1093/mnrasl/slv037}. Owing to the concentration of cosmic chronometer measurements at low redshift, the reconstruction is most reliable for $z \lesssim 1$. In this regime, $w(z)$ remains close to $-1$, consistent with a slowly evolving dark energy component. However, the uncertainty bands increase rapidly toward higher redshift due to data sparsity, and the GP reconstruction becomes less stable. In these regions, the mean reconstruction may temporarily enter the phantom regime ($w<-1$), but such behavior is not statistically significant given the large uncertainties, as shown in the top-left panel of Fig.~\ref{fig:w}.

The inclusion of DESI DR2 BAO measurements~\cite{DESI:2025zgx} significantly improves the reconstruction at intermediate redshifts. The combined CC-32 + DESI DR2 dataset reduces spurious fluctuations and tightens the uncertainty bands across $0 \lesssim z \lesssim 2$. The reconstructed equation of state exhibits clear deviations from $w=-1$ over a finite redshift range, with the cosmological constant value lying outside the $1\sigma$ confidence region in this interval. In particular, the reconstruction shows a crossing of the phantom divide around $z \sim 0.5$, indicating dynamical behavior of dark energy. These features arise from the strong constraints imposed by BAO measurements, which reveal structure in $w(z)$ that is not accessible using CC data alone. While the overall trend remains subject to uncertainties, the DESI DR2 data provide clear evidence for deviations from a strictly constant equation of state, as shown in the top-right panel of Fig.~\ref{fig:w}.

To further improve the low-redshift constraints, we incorporate Type Ia supernova observations using the Pantheon+ compilation~\cite{Scolnic:2017caz,Brout:2022vxf}. Owing to its large sample size and dense coverage at $z \lesssim 0.5$, the inclusion of Pantheon+ data significantly tightens the uncertainty bands and substantially improves the reconstruction in the low-redshift regime. The reconstructed equation of state remains close to the cosmological-constant limit at the present epoch, but the inferred value of $w_0$ exhibits a moderate ($\sim1.7$--$2.4\sigma$) preference for quintessence-like behaviour, depending on the assumed spatial curvature. This demonstrates that, although the Pantheon+ data strongly constrain the late-time expansion history, they do not uniquely favor a cosmological constant. Rather, they remain compatible with mild dynamical dark energy within the current observational uncertainties. As emphasized in recent DESI DR2 analyses, the inferred significance of dynamical dark energy depends sensitively on the adopted supernova compilation, and the present reconstruction exhibits a similar behaviour, as shown in the middle-left panel of Fig.~\ref{fig:w}.

We then extend the analysis by including CMB~\cite{Chen:2018dbv} information together with the CC-32, DESI DR2, and Pantheon+ datasets. The inclusion of the CMB produces only a mild impact on the reconstructed equation of state, indicating that the late time datasets already provide strong constraints on $w(z)$,as shown in the middle-right panel of Fig.~\ref{fig:w}. To assess the robustness of the reconstruction, we repeat the analysis by replacing the Pantheon+ compilation with the Union3~\cite{Rubin:2023jdq} and DES Y5~\cite{DES:2024jxu} supernova datasets, while keeping all other observational inputs unchanged, as shown in the bottom-left and bottom-right panels of Fig.~\ref{fig:w}, respectively. The reconstructed equation of state exhibits a qualitatively similar redshift evolution for all three supernova compilations, remaining close to the cosmological-constant limit throughout the observed redshift range. However, the statistical significance of the reconstructed present-day equation-of-state parameter depends on the adopted supernova compilation. The Union3 dataset remains broadly consistent with $\Lambda$CDM ($\sim1.1$--$1.5\sigma$), whereas the Pantheon+ and DES Y5 compilations show a moderate ($\sim2.1$--$2.7\sigma$) and stronger ($\sim3.1$--$3.3\sigma$) preference for quintessence-like behaviour, respectively. This trend is consistent with recent DESI DR2 analyses, which likewise find that the inferred significance of dynamical dark energy depends on the adopted supernova compilation\cite{DESI:2025zgx}.

Overall, the reconstructed equation of state remains close to the cosmological-constant limit, although the statistical significance of departures from $\Lambda$CDM depends on the adopted supernova compilation. While the Union3 dataset is broadly consistent with $\Lambda$CDM, the Pantheon+ and DES Y5 compilations exhibit increasing preference for quintessence-like behaviour. Small variations in the spatial curvature produce only minor changes in the reconstructed equation of state, indicating that these conclusions are robust against near-flat geometry.

\section{Discussion and Conclusion}
\label{sec:discussion}
In this work, we have presented a fully model-independent reconstruction of the late time expansion history and effective scalar field dynamics using GP. By combining cosmic chronometer measurements\cite{Zhang_2014, Stern:2010cv,Moresco:2012jh,Moresco_2016,10.1093/mnras/stx301,10.1093/mnrasl/slv037,Moresco:2020fbm}, DESI DR2 BAO data\cite{DESI:2025zgx}, Type Ia supernovae\cite{Scolnic:2017caz,Brout:2022vxf,Rubin:2023jdq,DES:2024jxu}, and CMB distance priors\cite{Chen:2018dbv}, we have obtained a consistent and robust description of dark energy evolution over the redshift range $0 \lesssim z \lesssim 2.5$.

A key outcome of this analysis is the complementary role of different datasets. Cosmic chronometers provide the primary constraints on the expansion rate, while DESI DR2 significantly improves the reconstruction at intermediate redshifts. Supernova data, particularly Pantheon+, tightly constrain the low-redshift regime through distance measurements. The inclusion of CMB shift parameters leads only to mild modifications, indicating that late time probes already provide strong constraints on the expansion history and that the overall reconstruction is internally consistent.

The reconstructed scalar field potential remains nearly constant at late times, with $V(0)/\rho_{c,0} \simeq 0.68\text{--}0.70$ (Fig.~\ref{fig:V}), consistent with the observed dark energy density. At the same time, the kinetic contribution (Fig.~\ref{fig:K}) remains subdominant across the entire redshift range, indicating slow-roll-like evolution. The reconstructed equation of state remains close to the cosmological-constant limit, with the inferred preference for dynamical dark energy depending strongly on the adopted supernova compilation. The Union3 data remain broadly consistent with $\Lambda$CDM at $\sim1.1$--$1.5\sigma$, while Pantheon+ and DES Y5 show moderate ($\sim2.1$--$2.7\sigma$) and stronger ($\sim3.1$--$3.3\sigma$) preferences for quintessence-like behaviour, respectively. Although mild deviations from $w=-1$, including a possible phantom crossing at intermediate redshifts, are reconstructed, their significance remains sensitive to the supernova compilation and current observational uncertainties, consistent with the compilation-dependent preference reported in recent DESI DR2 analyses~\cite{DESI:2025zgx}. The reconstruction of the potential slope parameter $|\lambda(z)|$ (Fig.~\ref{fig:lambda}) exhibits only mild evolution over the reconstructed redshift range, with present-day median values of approximately $0.8$--$1.0$, depending on the supernova compilation and spatial curvature. However, the associated uncertainties increase rapidly toward higher redshift because the reconstruction of $|\lambda(z)|$ requires higher-order derivatives of the expansion history. Consequently, current observations do not yet provide tight constraints on the slope of the effective scalar field potential. We also checked that the curvature parameter $\Gamma(z)$ is even more weakly constrained, as its reconstruction involves still higher-order derivatives [Eq.~(\ref{eq:Gamma})], leading to substantially larger propagated uncertainties.

We also find that spatial curvature introduces a mild degeneracy with the reconstructed dark energy dynamics, particularly in the observables discussed in Secs.~\ref{sec:V}, \ref{sec:K}, \ref{sec:lambda}, and \ref{sec:w}. As illustrated by the corresponding reconstruction figures, varying the spatial curvature within the range $\Omega_k=0$--$0.02$ produces only small shifts in the reconstructed quantities, which remain largely within their respective $1\sigma$ uncertainty bands. Thus, the main conclusions remain robust against small departures from spatial flatness. These results highlight the importance of jointly constraining spatial curvature and dark energy within a model independent framework. Overall, the reconstructed dynamics remain broadly consistent with a dark energy component close to the cosmological-constant scenario, while allowing for compilation-dependent evidence of mild dynamical evolution within current observational uncertainties. The GP framework therefore provides a powerful and flexible approach for probing the nature of dark energy from observational data without imposing a specific parametric model.
\bibliographystyle{apsrev4-2}
\bibliography{references}

@article{SupernovaSearchTeam:1998fmf,
    author = "Riess, Adam G. and others",
    collaboration = "Supernova Search Team",
    title = "{Observational evidence from supernovae for an accelerating universe and a cosmological constant}",
    eprint = "astro-ph/9805201",
    archivePrefix = "arXiv",
    doi = "10.1086/300499",
    journal = "Astron. J.",
    volume = "116",
    pages = "1009--1038",
    year = "1998"
}

@article{SupernovaCosmologyProject:1998vns,
    author = "Perlmutter, S. and others",
    collaboration = "Supernova Cosmology Project",
    title = "{Measurements of $\Omega$ and $\Lambda$ from 42 high redshift supernovae}",
    eprint = "astro-ph/9812133",
    archivePrefix = "arXiv",
    reportNumber = "LBNL-41801, LBL-41801",
    doi = "10.1086/307221",
    journal = "Astrophys. J.",
    volume = "517",
    pages = "565--586",
    year = "1999"
}

@article{Riess:2021jrx,
    author = "Riess, Adam G. and others",
    title = "{A Comprehensive Measurement of the Local Value of the Hubble Constant with 1 km s$^{-1}$ Mpc$^{-1}$ Uncertainty from the Hubble Space Telescope and the SH0ES Team}",
    eprint = "2112.04510",
    archivePrefix = "arXiv",
    primaryClass = "astro-ph.CO",
    doi = "10.3847/2041-$8213/ac5c5b",
    journal = "Astrophys. J. Lett.",
    volume = "934",
    number = "1",
    pages = "L7",
    year = "2022"
}

@article{Martin:2012bt,
    author = "Martin, Jerome",
    title = "{Everything You Always Wanted To Know About The Cosmological Constant Problem (But Were Afraid To Ask)}",
    eprint = "1205.3365",
    archivePrefix = "arXiv",
    primaryClass = "astro-ph.CO",
    doi = "10.1016/j.crhy.2012.04.008",
    journal = "Comptes Rendus Physique",
    volume = "13",
    pages = "566--665",
    year = "2012"
}

@article{Copeland:2006wr,
    author = "Copeland, Edmund J. and Sami, M. and Tsujikawa, Shinji",
    title = "{Dynamics of dark energy}",
    eprint = "hep-th/0603057",
    archivePrefix = "arXiv",
    doi = "10.1142/S021827180600942X",
    journal = "Int. J. Mod. Phys. D",
    volume = "15",
    pages = "1753--1936",
    year = "2006"
}

@article{Sahni:1999gb,
    author = "Sahni, Varun and Starobinsky, Alexei A.",
    title = "{The Case for a positive cosmological Lambda term}",
    eprint = "astro-ph/9904398",
    archivePrefix = "arXiv",
    reportNumber = "IUCAA-25-2000",
    doi = "10.1142/S0218271800000542",
    journal = "Int. J. Mod. Phys. D",
    volume = "9",
    pages = "373--444",
    year = "2000"
}

@article{Wetterich:1987fk,
    author = "Wetterich, C.",
    title = "{Cosmologies With Variable Newton's 'Constant'}",
    reportNumber = "Print-87-0757 (DESY), DESY-87-122",
    doi = "10.1016/0550-3213(88)90192-7",
    journal = "Nucl. Phys. B",
    volume = "302",
    pages = "645--667",
    year = "1988"
}

@article{Ratra:1987rm,
    author = "Ratra, Bharat and Peebles, P. J. E.",
    title = "{Cosmological Consequences of a Rolling Homogeneous Scalar Field}",
    reportNumber = "PUPT-1072",
    doi = "10.1103/PhysRevD.37.3406",
    journal = "Phys. Rev. D",
    volume = "37",
    pages = "3406",
    year = "1988"
}

@article{Wetterich:1987fm,
    author = "Wetterich, C.",
    title = "{Cosmology and the Fate of Dilatation Symmetry}",
    eprint = "1711.03844",
    archivePrefix = "arXiv",
    primaryClass = "hep-th",
    reportNumber = "PRINT-87-0756, DESY-87-123",
    doi = "10.1016/0550-3213(88)90193-9",
    journal = "Nucl. Phys. B",
    volume = "302",
    pages = "668--696",
    year = "1988"
}

@article{Planck:2013pxb,
    author = "Ade, P. A. R. and others",
    collaboration = "Planck",
    title = "{Planck 2013 results. XVI. Cosmological parameters}",
    eprint = "1303.5076",
    archivePrefix = "arXiv",
    primaryClass = "astro-ph.CO",
    reportNumber = "CERN-PH-TH-2013-129",
    doi = "10.1051/0004-6361/201321591",
    journal = "Astron. Astrophys.",
    volume = "571",
    pages = "A16",
    year = "2014"
}

@article{Copeland:1997et,
    author = "Copeland, Edmund J. and Liddle, Andrew R and Wands, David",
    title = "{Exponential potentials and cosmological scaling solutions}",
    eprint = "gr-qc/9711068",
    archivePrefix = "arXiv",
    reportNumber = "SUSX-TH-97-022, SUSSEX-AST-97-11-1, PU-RCG-97-20",
    doi = "10.1103/PhysRevD.57.4686",
    journal = "Phys. Rev. D",
    volume = "57",
    pages = "4686--4690",
    year = "1998"
}

@article{Steinhardt:1999nw,
    author = "Steinhardt, Paul J. and Wang, Li-Min and Zlatev, Ivaylo",
    title = "{Cosmological tracking solutions}",
    eprint = "astro-ph/9812313",
    archivePrefix = "arXiv",
    doi = "10.1103/PhysRevD.59.123504",
    journal = "Phys. Rev. D",
    volume = "59",
    pages = "123504",
    year = "1999"
}

@article{Bahamonde:2017ize,
    author = {Bahamonde, Sebastian and B\"ohmer, Christian G. and Carloni, Sante and Copeland, Edmund J. and Fang, Wei and Tamanini, Nicola},
    title = "{Dynamical systems applied to cosmology: dark energy and modified gravity}",
    eprint = "1712.03107",
    archivePrefix = "arXiv",
    primaryClass = "gr-qc",
    doi = "10.1016/j.physrep.2018.09.001",
    journal = "Phys. Rept.",
    volume = "775-777",
    pages = "1--122",
    year = "2018"
}

@article{Planck:2018vyg,
    author = "Aghanim, N. and others",
    collaboration = "Planck",
    title = "{Planck 2018 results. VI. Cosmological parameters}",
    eprint = "1807.06209",
    archivePrefix = "arXiv",
    primaryClass = "astro-ph.CO",
    doi = "10.1051/0004-6361/201833910",
    journal = "Astron. Astrophys.",
    volume = "641",
    pages = "A6",
    year = "2020",
    note = "[Erratum: Astron.Astrophys. 652, C4 (2021)]"
}

@article{Brout:2022vxf,
    author = "Brout, Dillon and others",
    title = "{The Pantheon+ Analysis: Cosmological Constraints}",
    eprint = "2202.04077",
    archivePrefix = "arXiv",
    primaryClass = "astro-ph.CO",
    doi = "10.3847/1538-4357/ac8e04",
    journal = "Astrophys. J.",
    volume = "938",
    number = "2",
    pages = "110",
    year = "2022"
}

@article{Scherrer:2007pu,
    author = "Scherrer, Robert J. and Sen, A. A.",
    title = "{Thawing quintessence with a nearly flat potential}",
    eprint = "0712.3450",
    archivePrefix = "arXiv",
    primaryClass = "astro-ph",
    doi = "10.1103/PhysRevD.77.083515",
    journal = "Phys. Rev. D",
    volume = "77",
    pages = "083515",
    year = "2008"
}

@article{Kamionkowski:2022pkx,
    author = "Kamionkowski, Marc and Riess, Adam G.",
    title = "{The Hubble Tension and Early Dark Energy}",
    eprint = "2211.04492",
    archivePrefix = "arXiv",
    primaryClass = "astro-ph.CO",
    doi = "10.1146/annurev-nucl-111422-024107",
    journal = "Ann. Rev. Nucl. Part. Sci.",
    volume = "73",
    pages = "153--180",
    year = "2023"
}

@article{Hossain:2023lxs,
    author = "Hossain, Md. Wali and Maqsood, Afaq",
    title = "{Comparison between axionlike and power law potentials in a cosmological background}",
    eprint = "2311.17825",
    archivePrefix = "arXiv",
    primaryClass = "astro-ph.CO",
    doi = "10.1103/PhysRevD.109.103512",
    journal = "Phys. Rev. D",
    volume = "109",
    number = "10",
    pages = "103512",
    year = "2024"
}

@article{10.1093/mnras/,
    author = {Ross, Ashley J. and Samushia, Lado and Howlett, Cullan and Percival, Will J. and Burden, Angela and Manera, Marc},
    title = "{The clustering of the SDSS DR7 main Galaxy sample – I. A 4 per cent distance measure at z = 0.15}",
    journal = {Monthly Notices of the Royal Astronomical Society},
    volume = {449},
    number = {1},
    pages = {835-847},
    year = {2015},
    month = {03},
    issn = {0035-8711},
    doi = {10.1093/mnras/stv154},
    url = {https://doi.org/10.1093/mnras/stv154},
    eprint = {https://academic.oup.com/mnras/article-pdf/449/1/835/13767551/stv154.pdf},
}

@article{eBOSS:2018cab,
    author = "Zarrouk, Pauline and others",
    collaboration = "eBOSS",
    title = "{The clustering of the SDSS-IV extended Baryon Oscillation Spectroscopic Survey DR14 quasar sample: measurement of the growth rate of structure from the anisotropic correlation function between redshift 0.8 and 2.2}",
    eprint = "1801.03062",
    archivePrefix = "arXiv",
    primaryClass = "astro-ph.CO",
    doi = "10.1093/mnras/sty506",
    journal = "Mon. Not. Roy. Astron. Soc.",
    volume = "477",
    number = "2",
    pages = "1639--1663",
    year = "2018"
}

@ARTICLE{2017MNRAS.470.2617A,
       author = {Shadab Alam et~al.},
        title = "{The clustering of galaxies in the completed SDSS-III Baryon Oscillation Spectroscopic Survey: cosmological analysis of the DR12 galaxy sample}",
      journal = {mnras},
         year = 2017,
        month = sep,
       volume = {470},
       number = {3},
        pages = {2617-2652},
          doi = {10.1093/mnras/stx721},
archivePrefix = {arXiv},
       eprint = {1607.03155},
 primaryClass = {astro-ph.CO},
       adsurl = {https://ui.adsabs.harvard.edu/abs/2017MNRAS.470.2617A}
}

@ARTICLE{2017A&A...608A.130D,
       author = {{du Mas des Bourboux} et~al.},
        title = "{Baryon acoustic oscillations from the complete SDSS-III Ly{\ensuremath{\alpha}}-quasar cross-correlation function at z = 2.4}",
      journal = {aap},
         year = 2017,
        month = dec,
       volume = {608},
          eid = {A130},
        pages = {A130},
          doi = {10.1051/0004-6361/201731731},
archivePrefix = {arXiv},
       eprint = {1708.02225},
 primaryClass = {astro-ph.CO},
       adsurl = {https://ui.adsabs.harvard.edu/abs/2017A&A...608A.130D}
}

@article{DESI:2024mwx,
    author = "Adame, A. G. and others",
    collaboration = "DESI",
    title = "{DESI 2024 VI: cosmological constraints from the measurements of baryon acoustic oscillations}",
    eprint = "2404.03002",
    archivePrefix = "arXiv",
    primaryClass = "astro-ph.CO",
    reportNumber = "FERMILAB-PUB-24-0154-PPD",
    doi = "10.1088/1475-7516/2025/02/021",
    journal = "JCAP",
    volume = "02",
    pages = "021",
    year = "2025"
}

@article{DESI:2024uvr,
    author = "Adame, A. G. and others",
    collaboration = "DESI",
    title = "{DESI 2024 III: baryon acoustic oscillations from galaxies and quasars}",
    eprint = "2404.03000",
    archivePrefix = "arXiv",
    primaryClass = "astro-ph.CO",
    reportNumber = "FERMILAB-PUB-24-0159-PPD",
    doi = "10.1088/1475-7516/2025/04/012",
    journal = "JCAP",
    volume = "04",
    pages = "012",
    year = "2025"
}

@article{DESI:2024lzq,
    author = "Adame, A. G. and others",
    collaboration = "DESI",
    title = "{DESI 2024 IV: Baryon Acoustic Oscillations from the Lyman alpha forest}",
    eprint = "2404.03001",
    archivePrefix = "arXiv",
    primaryClass = "astro-ph.CO",
    reportNumber = "FERMILAB-PUB-24-0147-PPD",
    doi = "10.1088/1475-7516/2025/01/124",
    journal = "JCAP",
    volume = "01",
    pages = "124",
    year = "2025"
}

@article{DESI:2024aqx,
    author = "Calderon, R. and others",
    collaboration = "DESI",
    title = "{DESI 2024: reconstructing dark energy using crossing statistics with DESI DR1 BAO data}",
    eprint = "2405.04216",
    archivePrefix = "arXiv",
    primaryClass = "astro-ph.CO",
    doi = "10.1088/1475-7516/2024/10/048",
    journal = "JCAP",
    volume = "10",
    pages = "048",
    year = "2024"
}

@article{Ramadan:2024kmn,
    author = "Ramadan, Omar F. and Sakstein, Jeremy and Rubin, David",
    title = "{DESI constraints on exponential quintessence}",
    eprint = "2405.18747",
    archivePrefix = "arXiv",
    primaryClass = "astro-ph.CO",
    doi = "10.1103/PhysRevD.110.L041303",
    journal = "Phys. Rev. D",
    volume = "110",
    number = "4",
    pages = "L041303",
    year = "2024"
}

@article{Jiang:2024xnu,
    author = "Jiang, Jun-Qian and Pedrotti, Davide and da Costa, Simony Santos and Vagnozzi, Sunny",
    title = "{Nonparametric late-time expansion history reconstruction and implications for the Hubble tension in light of recent DESI and type Ia supernovae data}",
    eprint = "2408.02365",
    archivePrefix = "arXiv",
    primaryClass = "astro-ph.CO",
    doi = "10.1103/PhysRevD.110.123519",
    journal = "Phys. Rev. D",
    volume = "110",
    number = "12",
    pages = "123519",
    year = "2024"
}

@article{Chevallier:2000qy,
    author = "Chevallier, Michel and Polarski, David",
    title = "{Accelerating universes with scaling dark matter}",
    eprint = "gr-qc/0009008",
    archivePrefix = "arXiv",
    doi = "10.1142/S0218271801000822",
    journal = "Int. J. Mod. Phys. D",
    volume = "10",
    pages = "213--224",
    year = "2001"
}

@article{Linder:2002et,
    author = "Linder, Eric V.",
    title = "{Exploring the expansion history of the universe}",
    eprint = "astro-ph/0208512",
    archivePrefix = "arXiv",
    doi = "10.1103/PhysRevLett.90.091301",
    journal = "Phys. Rev. Lett.",
    volume = "90",
    pages = "091301",
    year = "2003"
}

@article{Jassal:2005qc,
    author = "Jassal, Harvinder Kaur and Bagla, J. S. and Padmanabhan, T.",
    title = "{Observational constraints on low redshift evolution of dark energy: How consistent are different observations?}",
    eprint = "astro-ph/0506748",
    archivePrefix = "arXiv",
    doi = "10.1103/PhysRevD.72.103503",
    journal = "Phys. Rev. D",
    volume = "72",
    pages = "103503",
    year = "2005"
}

@article{Efstathiou:1999tm,
    author = "Efstathiou, G.",
    title = "{Constraining the equation of state of the universe from distant type Ia supernovae and cosmic microwave background anisotropies}",
    eprint = "astro-ph/9904356",
    archivePrefix = "arXiv",
    doi = "10.1046/j.1365-8711.1999.02997.x",
    journal = "Mon. Not. Roy. Astron. Soc.",
    volume = "310",
    pages = "842--850",
    year = "1999"
}

@article{DESI:2024kob,
    author = "Lodha, K. and others",
    collaboration = "DESI",
    title = "{DESI 2024: Constraints on physics-focused aspects of dark energy using DESI DR1 BAO data}",
    eprint = "2405.13588",
    archivePrefix = "arXiv",
    primaryClass = "astro-ph.CO",
    reportNumber = "FERMILAB-PUB-24-0756-PPD",
    doi = "10.1103/PhysRevD.111.023532",
    journal = "Phys. Rev. D",
    volume = "111",
    number = "2",
    pages = "023532",
    year = "2025"
}

@article{Sohail:2024oki,
    author = "Sohail, Sk. and Alam, Sonej and Akthar, Shiriny and Hossain, Md. Wali",
    title = "{Quintessential early dark energy}",
    eprint = "2408.03229",
    archivePrefix = "arXiv",
    primaryClass = "astro-ph.CO",
    doi = "10.1016/j.dark.2025.101948",
    journal = "Phys. Dark Univ.",
    volume = "48",
    pages = "101948",
    year = "2025"
}

@article{Akthar:2024tua,
    author = "Akthar, Shiriny and Hossain, Md. Wali",
    title = "{General parametrization for energy density of quintessence field}",
    eprint = "2411.15892",
    archivePrefix = "arXiv",
    primaryClass = "astro-ph.CO",
    doi = "10.1088/1475-7516/2025/04/024",
    journal = "JCAP",
    volume = "04",
    pages = "024",
    year = "2025"
}

@article{Colgain:2024mtg,
    author = "Colg\'ain, Eoin \'O. and Sheikh-Jabbari, M. M.",
    title = "{DESI and SNe: Dynamical Dark Energy, $\Omega_m$ Tension or Systematics?}",
    eprint = "2412.12905",
    archivePrefix = "arXiv",
    primaryClass = "astro-ph.CO",
    doi = "",
    journal = " ",
    volume = "",
    number = "",
    month = "12",
    year = "2024"
}

@article{Mukherjee:2024ryz,
    author = "Mukherjee, Purba and Sen, Anjan Ananda",
    title = "{Model-independent cosmological inference post DESI DR1 BAO measurements}",
    eprint = "2405.19178",
    archivePrefix = "arXiv",
    primaryClass = "astro-ph.CO",
    doi = "10.1103/PhysRevD.110.123502",
    journal = "Phys. Rev. D",
    volume = "110",
    number = "12",
    pages = "123502",
    year = "2024"
}

@article{DESI:2025zgx,
    author = "Abdul Karim, M. and others",
    collaboration = "DESI",
    title = "{DESI DR2 Results II: Measurements of Baryon Acoustic Oscillations and Cosmological Constraints}",
    eprint = "2503.14738",
    archivePrefix = "arXiv",
    primaryClass = "astro-ph.CO",
    reportNumber = "FERMILAB-PUB-25-0169-PPD",
    doi = "",
    journal = " ",
    volume = "",
    number = "",
    month = "3",
    year = "2025"
}

@article{Zhang_2014,
doi = {10.1088/1674-4527/14/10/002},
url = {https://doi.org/10.1088/1674-4527/14/10/002},
year = {2014},
month = {oct},
publisher = {},
volume = {14},
number = {10},
pages = {1221},
author = {Zhang Cong et al },
title = {Four new observational H(z) data from luminous red galaxies in the Sloan Digital Sky Survey data release seven},
journal = {Research in Astronomy and Astrophysics.}

}

@article{Stern:2010cv,
    author          = "Stern, Daniel and Jimenez, Raul and Verde, Licia and Kamionkowski, Marc and Stanford, S. Adam",
    title           = "{Cosmic chronometers: constraining the equation of state of dark energy. I: H(z) measurements}",
    eprint          = "0907.3149",
    archivePrefix   = "arXiv",
    primaryClass    = "astro-ph.CO",
    doi             = "10.1088/1475-7516/2010/02/008",
    journal         = "JCAP",
    volume          = "02",
    pages           = "008",
    year            = "2010"
}

@article{Moresco:2012jh,
    author          = "Moresco, Michele and Cimatti, Andrea and Jimenez, Raul and Pozzetti, Lucia and Zamorani, Gianfranco and others",
    title           = "{Improved constraints on the expansion rate of the Universe up to $z \sim 1.1$ from the spectroscopic evolution of cosmic chronometers}",
    eprint          = "1201.3609",
    archivePrefix   = "arXiv",
    primaryClass    = "astro-ph.CO",
    doi             = "10.1088/1475-7516/2012/08/006",
    journal         = "JCAP",
    volume          = "08",
    pages           = "006",
    year            = "2012"
}

@article{Moresco_2016,
doi = {10.1088/1475-7516/2016/05/014},
url = {https://doi.org/10.1088/1475-7516/2016/05/014},
year = {2016},
month = {may},
publisher = {},
volume = {2016},
number = {05},
pages = {014},
author = {Moresco, Michele and Pozzetti, Lucia and Cimatti, Andrea and Jimenez, Raul and Maraston, Claudia and Verde, Licia and Thomas, Daniel and Citro, Annalisa and Tojeiro, Rita and Wilkinson, David},
title = {A 6% measurement of the Hubble parameter at z∼0.45: direct evidence of the epoch of cosmic re-acceleration},
journal = {Journal of Cosmology and Astroparticle Physics.}

}

@article{10.1093/mnras/stx301,
    author = {Ratsimbazafy, A. L. and Loubser, S. I. and Crawford, S. M. and Cress, C. M. and Bassett, B. A. and Nichol, R. C. and Väisänen, P.},
    title = {Age-dating luminous red galaxies observed with the Southern African Large Telescope},
    journal = {Monthly Notices of the Royal Astronomical Society},
    volume = {467},
    number = {3},
    pages = {3239-3254},
    year = {2017},
    month = {02.}
    
}

@article{10.1093/mnrasl/slv037,
    author = {Moresco, Michele},
    title = {Raising the bar: new constraints on the Hubble parameter with cosmic chronometers at z$∼$2},
    journal = {Monthly Notices of the Royal Astronomical Society: Letters},
    volume = {450},
    number = {1},
    pages = {L16-L20},
    year = {2015},
    month = {04.}
    
}

@article{Krolewski:2025deb,
    author = "Krolewski, Alex and others",
    title = "{A measurement of $H_0$ from DESI DR1 using energy densities}",
    eprint = "2511.23432",
    archivePrefix = "arXiv",
    primaryClass = "astro-ph.CO",
    reportNumber = "FERMILAB-PUB-25-0870-PPD",
    doi = "10.5281/zenodo.17686137",
    journal = " ",
    volume = "",
    number = "",
    month = "11",
    year = "2025",
}

@article{Li:2025mmo,
    author = "Li, Zhizhong and Du, Shenshi and Liu, Tonghua and Zhu, Zonghong",
    title = "{Determination of H0 from current and forecasting time-delay lensed quasars at high redshift using cosmological model-independent method}",
    doi = "10.1142/S0218271825500609",
    journal = "Int. J. Mod. Phys. D",
    volume = "34",
    number = "13",
    pages = "2550060",
    year = "2025"


}

@article{Zlatev:1998tr,
    author = "Zlatev, Ivaylo and Wang, Li-Min and Steinhardt, Paul J.",
    title = "{Quintessence, cosmic coincidence, and the cosmological constant}",
    eprint = "astro-ph/9807002",
    archivePrefix = "arXiv",
    doi = "10.1103/PhysRevLett.82.896",
    journal = "Phys. Rev. Lett.",
    volume = "82",
    pages = "896--899",
    year = "1999"
}

@article{Capozziello:2025lor,
    author = "Capozziello, Salvatore and Chaudhary, Himanshu and Mustafa, G. and Pacif, S. K. J.",
    title = "{Evidence for Dynamical Dark Energy using DESI DR2 Ly$α$ Forest}",
    eprint = "2510.21976",
    archivePrefix = "arXiv",
    primaryClass = "astro-ph.CO",
    doi = "",
    journal = " ",
    volume = "",
    number = "",
    month = "10",
    year = "2025"
}

@article{Hossain:2025grx,
    author = "Hossain, Md. Wali and Maqsood, Afaq",
    title = "{Cosmological implications of tracker scalar fields: Testing the evidence for dynamical dark energy with recent data}",
    eprint = "2502.19274",
    archivePrefix = "arXiv",
    primaryClass = "astro-ph.CO",
    doi = "10.1103/cfwx-y336",
    journal = "Phys. Rev. D",
    volume = "112",
    number = "8",
    pages = "083504",
    year = "2025"
}

@article{Yao:2025wlx,
    author = "Yao, Zhibang and Ye, Gen and Silvestri, Alessandra",
    title = "{A general model for dark energy crossing the phantom divide}",
    eprint = "2508.01378",
    archivePrefix = "arXiv",
    primaryClass = "gr-qc",
    doi = "10.1088/1475-7516/2025/10/078",
    journal = "JCAP",
    volume = "10",
    pages = "078",
    year = "2025"
}

@article{Gialamas:2025pwv,
    author = {Gialamas, Ioannis D. and H{\"u}tsi, Gert and Raidal, Martti and Urrutia, Juan and Vasar, Martin and Veerm{\"a}e, Hardi},
    title = "{Quintessence and phantoms in light of DESI 2025}",
    eprint = "2506.21542",
    archivePrefix = "arXiv",
    primaryClass = "astro-ph.CO",
    doi = "10.1103/kdqc-y37v",
    journal = "Phys. Rev. D",
    volume = "112",
    number = "6",
    pages = "063551",
    year = "2025"
}

@article{Alam:2025epg,
    author = "Alam, Sonej and Hossain, Md. Wali",
    title = "{Beyond CPL: Evidence for dynamical dark energy in three-parameter models}",
    eprint = "2510.03779",
    archivePrefix = "arXiv",
    primaryClass = "astro-ph.CO",
    doi = "10.1088/1475-7516/2026/04/042",
    journal = "JCAP",
    volume = "04",
    pages = "042",
    year = "2026"
}

@article{Busti:2014aoa,
    author = "Busti, Vinicius C. and Clarkson, Chris and Seikel, Marina",
    editor = "Heavens, Alan and Starck, Jean-Luc and Krone-Martins, Alberto",
    title = "{The Value of $H_0$ from Gaussian Processes}",
    eprint = "1407.5227",
    archivePrefix = "arXiv",
    primaryClass = "astro-ph.CO",
    doi = "10.1017/S1743921314013751",
    journal = "IAU Symp.",
    volume = "306",
    pages = "25--27",
    year = "2014"
}

@article{Nair:2013sna,
    author = "Nair, Remya and Jhingan, Sanjay and Jain, Deepak",
    title = "{Exploring scalar field dynamics with Gaussian processes}",
    eprint = "1306.0606",
    archivePrefix = "arXiv",
    primaryClass = "astro-ph.CO",
    doi = "10.1088/1475-7516/2014/01/005",
    journal = "JCAP",
    volume = "01",
    pages = "005",
    year = "2014"
}

@book{Rasmussen:2005gp,
    author    = "Rasmussen, Carl Edward and Williams, Christopher K. I.",
    title     = "{Gaussian Processes for Machine Learning}",
    publisher = "MIT Press",
    address   = "Cambridge, MA",
    year      = "2005",
    isbn      = "9780262182539",
    series    = "Adaptive Computation and Machine Learning"
}

@article{Busti:2014dua,
    author = "Busti, Vinicius C. and Clarkson, Chris and Seikel, Marina",
    title = "{Evidence for a Lower Value for $H_0$ from Cosmic Chronometers Data?}",
    eprint = "1402.5429",
    archivePrefix = "arXiv",
    primaryClass = "astro-ph.CO",
    doi = "10.1093/mnrasl/slu035",
    journal = "Mon. Not. Roy. Astron. Soc.",
    volume = "441",
    pages = "11",
    year = "2014"
}

@article{Seikel:2012uu,
    author = "Seikel, Marina and Clarkson, Chris and Smith, Mathew",
    title = "{Reconstruction of dark energy and expansion dynamics using Gaussian processes}",
    eprint = "1204.2832",
    archivePrefix = "arXiv",
    primaryClass = "astro-ph.CO",
    doi = "10.1088/1475-7516/2012/06/036",
    journal = "JCAP",
    volume = "06",
    pages = "036",
    year = "2012"
}

@article{Seikel:2012cs,
    author = "Seikel, Marina and Yahya, Sahba and Maartens, Roy and Clarkson, Chris",
    title = "{Using H(z) data as a probe of the concordance model}",
    eprint = "1205.3431",
    archivePrefix = "arXiv",
    primaryClass = "astro-ph.CO",
    doi = "10.1103/PhysRevD.86.083001",
    journal = "Phys. Rev. D",
    volume = "86",
    pages = "083001",
    year = "2012"
}

@article{Seikel:2013fda,
    author = "Seikel, Marina and Clarkson, Chris",
    title = "{Optimising Gaussian processes for reconstructing dark energy dynamics from supernovae}",
    eprint = "1311.6678",
    archivePrefix = "arXiv",
    primaryClass = "astro-ph.CO",
    doi = "",
    journal = " ",
    volume = "",
    number = "",
    month = "11",
    year = "2013"
}

@article{Shafieloo:2012ht,
    author = "Shafieloo, Arman and Kim, Alex G. and Linder, Eric V.",
    title = "{Gaussian Process Cosmography}",
    eprint = "1204.2272",
    archivePrefix = "arXiv",
    primaryClass = "astro-ph.CO",
    doi = "10.1103/PhysRevD.85.123530",
    journal = "Phys. Rev. D",
    volume = "85",
    pages = "123530",
    year = "2012"
}

@article{Mukherjee:2022yyq,
    author = "Mukherjee, Purba and Levi Said, Jackson and Mifsud, Jurgen",
    title = "{Neural network reconstruction of H'(z) and its application in teleparallel gravity}",
    eprint = "2209.01113",
    archivePrefix = "arXiv",
    primaryClass = "astro-ph.CO",
    doi = "10.1088/1475-7516/2022/12/029",
    journal = "JCAP",
    volume = "12",
    pages = "029",
    year = "2022"
}

@article{Favale:2025mgk,
    author = "Favale, Arianna and G{\'o}mez-Valent, Adri{\`a} and Migliaccio, Marina",
    title = "{Revisiting model-independent constraints on spatial curvature and cosmic ladders calibration: updated and forecast analyses}",
    eprint = "2511.19332",
    archivePrefix = "arXiv",
    primaryClass = "astro-ph.CO",
    doi = "",
    journal = " ",
    volume = "",
    number = "",
    month = "11",
    year = "2025"
}

@article{Mukherjee:2020vkx,
    author = "Mukherjee, Purba and Banerjee, Narayan",
    title = "{Revisiting a non-parametric reconstruction of the deceleration parameter from combined background and the growth rate data}",
    eprint = "2007.15941",
    archivePrefix = "arXiv",
    primaryClass = "astro-ph.CO",
    doi = "10.1016/j.dark.2022.100998",
    journal = "Phys. Dark Univ.",
    volume = "36",
    pages = "100998",
    year = "2022"
}

@article{Ruchika:2025mkx,
    author = "Ruchika and Mukherjee, Purba and Favale, Arianna",
    title = "{Revisiting Gaussian Process Reconstruction for Cosmological Inference: The Generalised GP (Gen GP) Framework}",
    eprint = "2510.03742",
    archivePrefix = "arXiv",
    primaryClass = "astro-ph.CO",
    doi = "",
    journal = " ",
    volume = "",
    number = "",
    month = "10",
    year = "2025"
}

@article{Velazquez:2024aya,
    author = "Vel{\'a}zquez, Jos{\'e} de Jes{\'u}s and Escamilla, Luis A. and Mukherjee, Purba and V{\'a}zquez, J. Alberto",
    title = "{Non-Parametric Reconstruction of Cosmological Observables Using Gaussian Processes Regression}",
    eprint = "2410.02061",
    archivePrefix = "arXiv",
    primaryClass = "astro-ph.CO",
    doi = "10.3390/universe10120464",
    journal = "Universe",
    volume = "10",
    number = "12",
    pages = "464",
    year = "2024"
}

@article{Jiang:2025ilh,
    author = "Jiang, Jia-yan and Jiao, Kang and Zhang, Tong-Jie",
    title = "{Optimizing gaussian process kernels using nested sampling and ABC rejection for H(z) reconstruction}",
    eprint = "2506.21238",
    archivePrefix = "arXiv",
    primaryClass = "astro-ph.CO",
    doi = "10.1088/1475-7516/2025/11/065",
    journal = "JCAP",
    volume = "11",
    pages = "065",
    year = "2025"
}

@article{Holsclaw:2010sk,
    author = "Holsclaw, Tracy and Alam, Ujjaini and Sanso, Bruno and Lee, Herbert and Heitmann, Katrin and Habib, Salman and Higdon, David",
    title = "{Nonparametric Dark Energy Reconstruction from Supernova Data}",
    eprint = "1011.3079",
    archivePrefix = "arXiv",
    primaryClass = "astro-ph.CO",
    reportNumber = "LA-UR-09-07764",
    doi = "10.1103/PhysRevLett.105.241302",
    journal = "Phys. Rev. Lett.",
    volume = "105",
    pages = "241302",
    year = "2010"
}

@article{Holsclaw:2010nb,
    author = "Holsclaw, Tracy and Alam, Ujjaini and Sanso, Bruno and Lee, Herbert and Heitmann, Katrin and Habib, Salman and Higdon, David",
    title = "{Nonparametric Reconstruction of the Dark Energy Equation of State}",
    eprint = "1009.5443",
    archivePrefix = "arXiv",
    primaryClass = "astro-ph.CO",
    reportNumber = "LA-UR-09-05888",
    doi = "10.1103/PhysRevD.82.103502",
    journal = "Phys. Rev. D",
    volume = "82",
    pages = "103502",
    year = "2010"
}

@article{Holsclaw:2011wi,
    author = "Holsclaw, Tracy and Alam, Ujjaini and Sanso, Bruno and Lee, Herbie and Heitmann, Katrin and Habib, Salman and Higdon, David",
    title = "{Nonparametric Reconstruction of the Dark Energy Equation of State from Diverse Data Sets}",
    eprint = "1104.2041",
    archivePrefix = "arXiv",
    primaryClass = "astro-ph.CO",
    reportNumber = "LA-UR-11-01798",
    doi = "10.1103/PhysRevD.84.083501",
    journal = "Phys. Rev. D",
    volume = "84",
    pages = "083501",
    year = "2011"
}

@article{Favale:2023lnp,
    author = "Favale, Arianna and G{\'o}mez-Valent, Adri{\`a} and Migliaccio, Marina",
    title = "{Cosmic chronometers to calibrate the ladders and measure the curvature of the Universe. A model-independent study}",
    eprint = "2301.09591",
    archivePrefix = "arXiv",
    primaryClass = "astro-ph.CO",
    doi = "10.1093/mnras/stad1621",
    journal = "Mon. Not. Roy. Astron. Soc.",
    volume = "523",
    number = "3",
    pages = "3406--3422",
    year = "2023"
}

@article{Scolnic:2017caz,
    author         = "Scolnic, D. M. and others",
    title          = "{The Complete Light-curve Sample of Spectroscopically Confirmed SNe Ia from Pan-STARRS1 and Cosmological Constraints from the Combined Pantheon Sample}",
    journal        = "Astrophys. J.",
    volume         = "859",
    year           = "2018",
    number         = "2",
    pages          = "101",
    doi            = "10.3847/1538-4357/aab9bb",
    eprint         = "1710.00845",
    archivePrefix  = "arXiv",
    primaryClass   = "astro-ph.CO"
}

@article{Johnson:2025blf,
    author = "Johnson, Joseph P. and Jassal, H. K.",
    title = "{Kernel dependence of the Gaussian process reconstruction of late Universe expansion history}",
    eprint = "2503.04273",
    archivePrefix = "arXiv",
    primaryClass = "astro-ph.CO",
    doi = "10.1140/epjc/s10052-025-14732-7",
    journal = "Eur. Phys. J. C",
    volume = "85",
    number = "9",
    pages = "996",
    year = "2025"
}

@article{Ormondroyd:2025exu,
    author = "Ormondroyd, A. N. and Handley, W. J. and Hobson, M. P. and Lasenby, A. N.",
    title = "{Non-parametric reconstructions of dynamical dark energy via flexknots}",
    eprint = "2503.08658",
    archivePrefix = "arXiv",
    primaryClass = "astro-ph.CO",
    doi = "10.1093/mnras/staf1144",
    journal = "Mon. Not. Roy. Astron. Soc.",
    volume = "541",
    number = "4",
    pages = "3388--3400",
    year = "2025"
}

@article{Sangwan:2017kxi,
    author = "Sangwan, Archana and Mukherjee, Ankan and Jassal, H. K.",
    title = "{Reconstructing the dark energy potential}",
    eprint = "1712.05143",
    archivePrefix = "arXiv",
    primaryClass = "astro-ph.CO",
    doi = "10.1088/1475-7516/2018/01/018",
    journal = "JCAP",
    volume = "01",
    pages = "018",
    year = "2018"
}

@article{Moresco:2020fbm,
    author = "Moresco, Michele and Jimenez, Raul and Verde, Licia and Cimatti, Andrea and Pozzetti, Lucia",
    title = "{Setting the Stage for Cosmic Chronometers. II. Impact of Stellar Population Synthesis Models Systematics and Full Covariance Matrix}",
    eprint = "2003.07362",
    archivePrefix = "arXiv",
    primaryClass = "astro-ph.GA",
    doi = "10.3847/1538-4357/ab9eb0",
    journal = "Astrophys. J.",
    volume = "898",
    number = "1",
    pages = "82",
    year = "2020"
}

@article{Chen:2018dbv,
    author = "Chen, Lu and Huang, Qing-Guo and Wang, Ke",
    title = "{Distance Priors from Planck Final Release}",
    eprint = "1808.05724",
    archivePrefix = "arXiv",
    primaryClass = "astro-ph.CO",
    doi = "10.1088/1475-7516/2019/02/028",
    journal = "JCAP",
    volume = "02",
    pages = "028",
    year = "2019"
}

@article{Shlivko:2026jxa,
    author = "Shlivko, David and Poulin, Vivian",
    title = "{Phantom-Crossing Dark Energy and the $Ω_m$ Tug-of-War}",
    eprint = "2603.22406",
    archivePrefix = "arXiv",
    primaryClass = "astro-ph.CO",
    doi = "",
    journal = " ",
    volume = "",
    number = "",
    month = "3",
    year = "2026"
}

@article{Gokcen:2026pkq,
    author = {G{\"o}k{\c{c}}en, Mine and Akarsu, {\"O}zg{\"u}r and Di Valentino, Eleonora},
    title = "{Revisiting CPL with sign-switching density: To cross or not to cross the NECB}",
    eprint = "2602.21169",
    archivePrefix = "arXiv",
    primaryClass = "astro-ph.CO",
    doi = "10.1016/j.dark.2026.102273",
    journal = "Phys. Dark Univ.",
    volume = "52",
    pages = "102273",
    year = "2026"
}

@article{Li:2025ops,
    author = "Li, Jun-Xian and Wang, Shuang",
    title = "{Reconstructing dark energy with model independent methods after DESI DR2}",
    eprint = "2506.22953",
    archivePrefix = "arXiv",
    primaryClass = "astro-ph.CO",
    doi = "10.1140/epjc/s10052-025-15065-1",
    journal = "Eur. Phys. J. C",
    volume = "85",
    number = "11",
    pages = "1308",
    year = "2025"
}

@article{Zhang:2025bmk,
    author = "Zhang, Xue and Xu, Yin-Hao and Sang, Yu",
    title = "{Reconstruction of dark energy using DESI DR2}",
    eprint = "2511.02220",
    archivePrefix = "arXiv",
    primaryClass = "astro-ph.CO",
    doi = "10.1088/1572-9494/ae1a5b",
    journal = "Commun. Theor. Phys.",
    volume = "78",
    number = "3",
    pages = "035404",
    year = "2026"
}

@article{DES:2024jxu,
    author = "Abbott, T. M. C. and others",
    collaboration = "DES",
    title = "{The Dark Energy Survey: Cosmology Results with {\ensuremath{\sim}}1500 New High-redshift Type Ia Supernovae Using the Full 5 yr Data Set}",
    eprint = "2401.02929",
    archivePrefix = "arXiv",
    primaryClass = "astro-ph.CO",
    reportNumber = "FERMILAB-PUB-23-0821-PPD, DES-2023-805",
    doi = "10.3847/2041-8213/ad6f9f",
    journal = "Astrophys. J. Lett.",
    volume = "973",
    number = "1",
    pages = "L14",
    year = "2024"
}

@article{Rubin:2023jdq,
    author = "Rubin, David and others",
    title = "{Union Through UNITY: Cosmology with 2,000 SNe Using a Unified Bayesian Framework}",
    eprint = "2311.12098",
    archivePrefix = "arXiv",
    primaryClass = "astro-ph.CO",
    doi = "10.3847/1538-4357/adc0a5",
    journal = "Astrophys. J.",
    volume = "986",
    number = "2",
    pages = "231",
    year = "2025"
}

@article{Koksbang:2026wvh,
    author = "Koksbang, S. M. and Heinesen, A.",
    title = "{Model-independent constraints on generalized FLRW consistency relations with bootstrap-based symbolic regression}",
    eprint = "2604.05822",
    archivePrefix = "arXiv",
    primaryClass = "astro-ph.CO",
    doi = "",
    journal = " ",
    volume = "",
    number = "",
    month = "4",
    year = "2026"
}

@article{MAQSOOD2026140456,
title = {Derivative Hierarchy as the Origin of Kernel-Dependent Trends in Gaussian Process Reconstructions of the Hubble Parameter},
journal = {Physics Letters B},
pages = {140456},
year = {2026},
issn = {0370-2693},
doi = {https://doi.org/10.1016/j.physletb.2026.140456},
url = {https://www.sciencedirect.com/science/article/pii/S0370269326003096},
author = {Afaq Maqsood}

}

@article{Li:2026hwq,
    author = "Li, Xiaolei and Liu, Tonghua and Li, Tian-Nuo and Du, Guo-Hong and Shafieloo, Arman and Biesiada, Marek",
    title = "{Metastability in Emergent Dark Energy: A New Framework Confronting Cosmological Observations}",
    doi = "10.3847/2041-8213/ae5a36",
    journal = "Astrophys. J. Lett.",
    volume = "1001",
    number = "1",
    pages = "L21",
    year = "2026"
}

@article{You:2025uon,
    author = "You, Changyu and Wang, Dan and Yang, Tao",
    title = "{Dynamical dark energy implies a coupled dark sector: Insights from DESI DR2 via a data-driven approach}",
    eprint = "2504.00985",
    archivePrefix = "arXiv",
    primaryClass = "astro-ph.CO",
    doi = "10.1103/f6v7-n9fr",
    journal = "Phys. Rev. D",
    volume = "112",
    number = "4",
    pages = "043503",
    year = "2025"
}

@article{Ormondroyd:2025iaf,
    author = "Ormondroyd, A. N. and Handley, W. J. and Hobson, M. P. and Lasenby, A. N.",
    title = "{Comparison of dynamical dark energy with {\ensuremath{\Lambda}}CDM in light of DESI DR2}",
    eprint = "2503.17342",
    archivePrefix = "arXiv",
    primaryClass = "astro-ph.CO",
    doi = "",
    journal = " ",
    volume = "",
    number = "",
    month = "3",
    year = "2025"
}

@article{Jesus:2021bxq,
    author = "Jesus, J. F. and Valentim, R. and Escobal, A. A. and Pereira, S. H. and Benndorf, D.",
    title = "{Gaussian processes reconstruction of the dark energy potential}",
    eprint = "2112.09722",
    archivePrefix = "arXiv",
    primaryClass = "astro-ph.CO",
    doi = "10.1088/1475-7516/2022/11/037",
    journal = "JCAP",
    volume = "11",
    pages = "037",
    year = "2022"
}

@article{Niu:2023hak,
    author = "Niu, Jing and Jiao, Kang and He, Peng and Zhang, Tong-Jie",
    title = "{Reconstruction of the Dark Energy Scalar Field Potential by Gaussian Process}",
    eprint = "2305.04752",
    archivePrefix = "arXiv",
    primaryClass = "astro-ph.CO",
    doi = "10.3847/1538-4357/ad5fef",
    journal = "Astrophys. J.",
    volume = "972",
    number = "1",
    pages = "14",
    year = "2024"
}

@article{Jesus:2022xwb,
    author = "Jesus, J. F. and Benndorf, D. and Escobal, A. A. and Pereira, S. H.",
    title = "{From Hubble to snap parameters: a Gaussian process reconstruction}",
    eprint = "2212.12346",
    archivePrefix = "arXiv",
    primaryClass = "astro-ph.CO",
    doi = "10.1093/mnras/stae120",
    journal = "Mon. Not. Roy. Astron. Soc.",
    volume = "528",
    number = "2",
    pages = "1573--1581",
    year = "2024"
}

@article{Liu:2023agr,
    author = "Liu, Jinyi and Qiao, Ling and Chang, Baorong and Xu, Lixin",
    title = "{Revisiting cosmography via Gaussian process}",
    doi = "10.1140/epjc/s10052-023-11545-4",
    journal = "Eur. Phys. J. C",
    volume = "83",
    number = "5",
    pages = "374",
    year = "2023"
}

@article{DESI:2025wyn,
    author = "Gu, Gan and others",
    collaboration = "DESI",
    title = "{Dynamical dark energy in light of the DESI DR2 baryonic acoustic oscillations measurements}",
    eprint = "2504.06118",
    archivePrefix = "arXiv",
    primaryClass = "astro-ph.CO",
    reportNumber = "FERMILAB-PUB-25-0235-PPD",
    doi = "10.1038/s41550-025-02669-6",
    journal = "Nature Astron.",
    volume = "9",
    number = "12",
    pages = "1879--1889",
    year = "2025",
    note = "[Erratum: Nature Astron. 9, 1898 (2025)]"
}

@article{DESI:2025fii,
    author = "Lodha, K. and others",
    collaboration = "DESI",
    title = "{Extended dark energy analysis using DESI DR2 BAO measurements}",
    eprint = "2503.14743",
    archivePrefix = "arXiv",
    primaryClass = "astro-ph.CO",
    reportNumber = "FERMILAB-PUB-25-0164-PPD",
    doi = "10.1103/w4c6-1r5j",
    journal = "Phys. Rev. D",
    volume = "112",
    number = "8",
    pages = "083511",
    year = "2025"
}

@article{GuptaChoudhury:2026gsl,
    author = "Gupta Choudhury, Shibendu and Mukherjee, Purba and Di Valentino, Eleonora and Sen, Anjan A.",
    title = "{Model-Independent Indication for a Localized Anomaly in the Late-Time Expansion History}",
    eprint = "2607.13009",
    archivePrefix = "arXiv",
    primaryClass = "astro-ph.CO",
    doi = "",
    journal = " ",
    volume = "",
    number = "",
    month = "7",
    year = "2026"
}

@book{Gelman2013,
  author    = {Andrew Gelman and John B. Carlin and Hal S. Stern and
               David B. Dunson and Aki Vehtari and Donald B. Rubin},
  title     = {Bayesian Data Analysis},
  edition   = {3},
  year      = {2013},
  publisher = {Chapman and Hall/CRC},
  address   = {Boca Raton, FL},
  isbn      = {9781439898208}
}

@article{Barua:2025jid,
    author = "Barua, Shubham and Desai, Shantanu and Lopez-Hernandez, Mauricio and Colg{\'a}in, Eoin {\'O}.",
    title = "{On frequentist confidence intervals in a non-Gaussian regime}",
    eprint = "2508.10633",
    archivePrefix = "arXiv",
    primaryClass = "astro-ph.CO",
    doi = "10.1140/epjc/s10052-025-15250-2",
    journal = "Eur. Phys. J. C",
    volume = "86",
    number = "1",
    pages = "47",
    year = "2026"
}

@article{Maqsood:2026krg,
    author = "Maqsood, Afaq and Duary, Tanima",
    title = "{Model-independent reconstruction of cosmic thermodynamics and dark energy dynamics}",
    eprint = "2604.18723",
    archivePrefix = "arXiv",
    primaryClass = "astro-ph.CO",
    doi = "10.1140/epjc/s10052-026-16223-9",
    journal = "Eur. Phys. J. C",
    volume = "86",
    number = "8",
    pages = "963",
    year = "2026"
}

\end{document}